%% file: main.tex
\documentclass[12pt]{article}
\usepackage{amsmath}
\usepackage{amssymb,amsfonts}
\usepackage{amsthm}
\usepackage[all]{xy}
\usepackage{graphicx}
\usepackage{setspace}
\usepackage{hyperref}
\usepackage{tikz-cd}
\usepackage{verbatim}
\usepackage{multirow}
\usepackage{slashed}

\usepackage{tensor}
\usepackage{braket}
\usepackage{mathtools}

\usepackage[parfill]{parskip}
\usepackage{tabu}
\usepackage{float}
\usepackage[utf8]{inputenc}

\usepackage{caption}
\usepackage{subcaption}
\usepackage{stackengine}

\usepackage{soul}
\usepackage{comment}
\usepackage{mathtools}
\usepackage{stmaryrd}

\usepackage{diagbox}
\definecolor{myblue}{rgb}{0,0,0.8}

\tikzcdset{scale cd/.style={every label/.append style={scale=#1},
    cells={nodes={scale=#1}}}}

\numberwithin{equation}{section}
\numberwithin{table}{section}

\newcommand{\be}{\begin{equation}}
\newcommand{\ee}{\end{equation}}
\def\beq{\begin{eqnarray}}
\def\eeq{\end{eqnarray}}
\def\ba{\begin{eqnarray}}
\def\ea{\end{eqnarray}}

\def\ep1{\epsilon_1}
\def\eps2{\epsilon_2}
\def\t1{\theta_1}

\newcommand{\IZ}{\mathbb{Z}}
\newcommand{\IC}{\mathbb{C}}
\newcommand{\IP}{\mathbb{P}}

\newcommand{\IH}{\mathbb{H}}

\newcommand{\IF}{\mathbb{F}}

\usepackage{marginnote}

\newcommand{\tr}{\mathrm{Tr\, }}

\newcommand{\nn}{\nonumber}

\newcommand{\cW}{{\cal W}}
\newcommand{\cN}{{\cal N}}
\newcommand{\cM}{{\cal M}}

\newcommand{\cH}{{\cal H}}

\newcommand{\cO}{{\cal O}}

\newcommand{\LambdaW}{\Lambda_{w}}

\newcommand{\cha}{\mathrm{ch}\,}

\newcommand{\ER}{E_{\mathrm{R}}}
\newcommand{\E}{E_{\mathrm{R}}}
\newcommand{\ENS}{E_{\mathrm{NS}}}
\newcommand{\fpq}{x} 
\newcommand{\lpq}{\fpq_0}
\newcommand{\tM}{\widetilde \cM}
\newcommand{\casimirNS}{E_{\mathrm{NS}}^0}
\newcommand{\casimirR}{E_{\mathrm{R}}^0}
\newcommand{\Fff}{(F_4)_4}
\newcommand{\dominant}{\Lambda^+}
\newcommand{\affinedominant}{\hat \Lambda^+}
\newcommand{\dominantflavor}{\affinedominant(F)_{k_F}}
\newcommand{\dominantgauge}{\dominant(G)_{-n}}
\newcommand{\affineg}{\hat{\mathfrak{g}}}
\newcommand{\affineWeyl}{\hat{\cW}}
\newcommand{\finiteWeyl}{\cW}

\begin{document}
	\begin{titlepage}
		{}~ \hfill\vbox{ \hbox{} }\break

		\vskip 1 cm

		\begin{center}
			\Large \bf Affine characters at negative level and elliptic genera of non-critical strings
		\end{center}
		
		\vskip 0.8 cm
		\centerline{David Jaramillo Duque and Amir-Kian Kashani-Poor }

		\vskip 0.2in
		
		\begin{center}{\footnotesize
				\begin{tabular}{c}
					{\em LPENS, CNRS, PSL Research University, Sorbonne Universit\'{e}s, UPMC, 75005 Paris, France}\\[0ex]
				\end{tabular}
		}\end{center}

		\setcounter{footnote}{0}
		\renewcommand{\thefootnote}{\arabic{footnote}}
		\vskip 60pt
		\begin{abstract} 
	       We study the elliptic genera of the non-critical strings of six dimensional superconformal field theories from the point of view of the strings' worldsheet theory. We formulate a general ansatz for these in terms of characters of the affine Lie algebra associated to the 6d gauge group at negative level, and provide ample evidence for the validity of this ansatz for 6d theories obtained via F-theory compactification on elliptically fibered Calabi-Yau manifolds over a Hirzebruch base. We obtain novel closed form results for many elliptic genera in terms of our ansatz, and show that our results specialize consistently when moving along Higgsing trees.
        
		\end{abstract}
		
		{\let\thefootnote\relax
			\footnotetext{\tiny david.jaramillo-duque@phys.ens.fr,  amir-kian.kashani-poor@phys.ens.fr}}
		
	\end{titlepage}

	\pagebreak
	
	\vfill \eject

	\tableofcontents
	\newpage

    \section{Introduction}
    The elliptic genus is a powerful tool to study two dimensional superconformal field theories for which an explicit description is not known. Due to its cohomological nature, it is robust under continuous deformations of the theory which do not change the asymptotics of field space. As such, it is invariant under renormalization group flow and can be computed if a Lagrangian UV description of the theory is available. Such descriptions exist e.g. for the (2,2) theories which arise upon considering Calabi-Yau compactifications of string theory \cite{Witten:1993yc} or for a small sample of the (0,2) theories which will be the object of this paper. Given such a description, it is possible to obtain exact results for the elliptic genus in terms of Jacobi $\theta$-functions and the Dedekind $\eta$-function \cite{Benini:2013a,Benini:2013xpa,Kim:2014dza,DelZotto:2018tcj,Haghighat:2014vxa,Kim:2018gjo,Kim:2016foj,Kim:2015fxa,Yun_2016,Gadde:2015tra,Haghighat:2013,Haghighat:2013tka,Hohenegger:2013ala,Putrov:2015jpa,Iqbal:2007ii,Agarwal:2018ejn}. While this solves the problem of computing the elliptic genus, the symmetries of the conformal fixed point of the theory will typically not be manifest. The scarceness of theories with available UV descriptions, and the obfuscation of the IR physics in the $\theta$-function form of the elliptic genus, motivate this work.
    
    We will study the elliptic genera of the non-critical strings of 6d superconformal field theories obtained via F-theory compactification on Calabi-Yau threefolds elliptically fibered over Hirzebruch surfaces $\IF_n$. We denote such theories as $G_n$, with $G$ specifying the 6d gauge symmetry.\footnote{With a very few exceptions at $n=1,2$, this specifies all theories in the class we are considering uniquely. For the exceptions, the flavor group $F$ must also be indicated.} The authors of \cite{DelZotto:2018tcj} propose that these genera take the universal form
    \begin{equation} \label{eq:affineERintro}
        \E = \sum_{\lambda \in \dominantflavor} \hat{\chi}^F_\lambda(m_F,q) \, \xi^{n,G}_\lambda(m_G,q,v)
    \end{equation}
    with $\hat{\chi}_\lambda^F$ indicating the affine character of the integrable representation of the flavor group $F$ with highest affine dominant weight $\lambda$ at level $k_F$ -- we denote the set of such weights as $\dominantflavor$. Extending their work, we conjecture that $\xi^{n,G}_\lambda$ satisfies the ansatz 
    \begin{equation} \label{eq:affineAnsatzIntro}
    \xi_{\lambda}^{n,G}=\frac{1}{\eta(q)}\sum_{\omega \in \dominantgauge} \hat\chi_\omega(m_G,q) \sum_{k\in \mathbb Z} c^\lambda_{\omega,k} \,  q^{-\frac{k^2}{4\kappa}}v^k \,.
    \end{equation}
    with 
    \begin{equation} \label{eq:constraintIntro}
        c^\lambda_{\omega,k}\in \{0,\pm 1,\pm 2\} \,, \quad \#\{k \in \IZ\, | \, c_{\omega, k}^\lambda \neq 0 \} < \infty \,,
     \end{equation}
     i.e.,  the coefficients $c_{\omega, k}^\lambda$ are non-vanishing at given $\omega$ and $\lambda$ for only a finite number of integers $k$. A surprise, already pointed out in \cite{DelZotto:2016pvm,DelZotto:2018tcj}, is the occurrence of non-integrable representations of the affine Lie algebra associated to $G$ at negative level $-n$. The notation $\dominant(G)_{k}$ is meant to signify the set of affine weights at level $k$ with dominant finite part. Unlike $\affinedominant(F)_{k}$, this set is infinite. Such representations have only occurred sporadically in the physics literature, e.g. in \cite{Beem:2013sza, Dedushenko:2015opz,Dedushenko:2017osi}.\footnote{See \cite{Eager:2019zrc} for the occurrence of even more general non-integrable representations.} Their relevance for the elliptic genus of a handful of theories for which a UV description was known, namely $(C_r)_1$, $r=0,1,2$, $(D_4)_4$ and  $(G_2)_3$, was shown in \cite{DelZotto:2018tcj}. In this work, we demonstrate that the elliptic genera for a large number of theories for which no UV description is known can be expressed in terms of affine characters in the form \eqref{eq:affineERintro} with $\xi_\lambda^{n,G}$ given by our ansatz \eqref{eq:affineAnsatzIntro}, satisfying the constraints \eqref{eq:constraintIntro}. To be precise, we are able to compute elliptic genera for these theories up to some order in $q$ and $v$ by combining known constraints on the elliptic genus with the new constraints coming from our ansatz. The obtained results pass numerous consistency tests, leaving little doubt regarding their validity. 

    Our results can be used to provide boundary conditions for a universal ansatz for the elliptic genus in terms of Weyl invariant Jacobi forms \cite{Huang:2015sta, DelZotto:2016pvm, Gu:2017ccq, DelZotto:2017mee,Kim:2018gak} (referred to as the modular ansatz henceforth), leading to an exact expression for the genus. Mainly, though, we are interested in what they reveal about the nature of the worldsheet theories of the non-critical strings. The list of representations $\omega$ contributing to the elliptic genus, as well as the explicit form of the polynomials in $q$ and $v$ multiplying the characters $\hat \chi_\omega$ in equation \eqref{eq:affineAnsatzIntro} which we determine furnish non-trivial information regarding the structure of these theories. For the moment however, the nature of the non-integrable affine symmetry remains to be understood.

    Our original motivation for pursuing the calculations presented in this paper arose from our study in \cite{Duan:2020imo} of Weyl symmetry enhancement from $\cW_G$ to $\cW$ along  Higgsing trees as detected by the modular ansatz for the elliptic genus in terms of Jacobi forms. It is natural to ask whether the elliptic genus of a theory experiencing such an enhancement can beneficially be expanded in terms of affine characters $\hat \chi_\omega$ of the larger Lie algebra exhibiting $\cW$ as its Weyl symmetry. We will explain why this is always possible, and, sadly, why it does not appear to be beneficial, i.e. to lead to a simplification of the polynomial contribution in the ansatz \eqref{eq:affineAnsatzIntro} multiplying $\hat \chi_\omega(m_G,Q)$.
    
    This paper is organized as follows: section \ref{s:theories_considered} offers a quick review of the 6d spacetime theories under considerations and collects known and lesser known facts about the non-critical string in their spectrum and their worldsheet theory. We carefully introduce the $U(1)_v$ symmetry which will play an important role in our analysis. We introduce the constraints on the elliptic genera which will allow us to compute it in section \ref{s:four_constraints}. Some of the constraints that we wish to impose are on the elliptic genus with NS-NS boundary conditions. Section \ref{s:four_constraints} therefore also includes a subsection \ref{ss:intermezzo} in which we review and discuss the map introduced in \cite{DelZotto:2018tcj} between the R-R and the NS-NS elliptic genus. Section \ref{sec:algorithm} introduces the strategy to compute elliptic genera by imposing the constraints introduced in section \ref{s:four_constraints}. Much of the logic here follows \cite{DelZotto:2018tcj}, with the crucial addition of our constraints on the form of the function $\xi_\lambda^{n,G}$. In section \ref{s:closed_form_results}, we conjecture closed form results for the elliptic genera of multiple theories in terms of the ansatz \eqref{eq:affineERintro} and \eqref{eq:affineAnsatzIntro}. Section \ref{s:Higgs_tree_relations} discusses how, for multiple pairs of theories related by Higgsing, the elliptic genus of the parent theory specializes to the elliptic genus of the daughter theory. We also discuss the negative result regarding the expansion of the elliptic genus in terms of larger affine characters mentioned above. We conclude by summarizing the observations in this paper which would merit being understood purely from the point of view of the conformal field theory of the worldsheet theory of the non-critical string. Several appendices complete the paper. We review well-known facts regarding elliptic genera in appendix \ref{a:elliptic_genera}. Appendix \ref{a:Kazhdan-Lusztig} summarizes how we computed characters of negative level representations, an important technical ingredient in our calculations. Appendix \ref{a:polynomials} presents our results for the affine character expansion of the elliptic genus of a large number of theories. We end in appendix \ref{a:higgsing_trees} with a summary of the theories in the class we are considering, together with comments regarding the status of their elliptic genus. A file containing all negative level affine characters underlying the computations in this paper can be downloaded from \url{http://www.phys.ens.fr/~kashani/}.

    \section{6d theories and their non-critical strings} \label{s:theories_considered}
    \subsection{Spacetime theory} \label{ss:spacetime_theory}
    
    The theories we will consider in this paper are six dimensional rank 1 superconformal field theories with $(1,0)$ supersymmetry \cite{Gaiotto:2014lca,DelZotto:2014hpa,Bhardwaj:2015oru,DelZotto:2017pti,DelZotto:2018tcj,DelZotto:2016pvm,Losev:2003py,Lee:2018urn,Heckman:2018jxk}.\footnote{The rank indicates the number of tensor multiplets in the theory. $(1,0)$ is the minimal amount of supersymmetry possible in 6d: 8 real supercharges.}
    The existence of such theories as non-trivial infrared fixed points of 6d quantum field theories was one of the surprises that arose from the construction of 6d theories within the framework of string theory in the nineties \cite{Vafa:1996xn,Schwarz:1995dk}: the authors of \cite{Seiberg:1996vs} argued that singularities that occur in the moduli space of such theories must be due to IR dynamics, and concluded that the latter can therefore not be trivial.

    The theories we will consider can be constructed within F-theory by compactification on elliptically fibered Calabi-Yau manifolds $X$ over Hirzebruch bases $B=\IF_n$ \cite{Morrison:1996na,Morrison:1996pp,Morrison:2012np,Heckman:2013pva,Heckman:2015bfa}. These surfaces can be presented as $\IP^1$ fibrations over $\IP^1$, hence give rise to 6d supergravity theories with two tensor multiplets. Decompactifying the fiber $\IP^1$ (i.e. replacing $\IF_n$ by $\cO(-n) \rightarrow \IP^1$) decouples gravity and yields rank 1 theories with superconformal fixed points. 
    
    The base $\IF_n$ typically does not determine the elliptically fibered Calabi-Yau manifold above it uniquely, but determines a sometimes branched sequence of geometries called Higgsing trees. At the root of each tree lies the least singular fibration the base permits, leading to what is called the maximally Higgsed theory. Moving away from the root, one encounters increasingly singular fibrations, leading to theories with larger gauge symmetry and typically also larger charged matter content. Throughout this paper, we will refer to the gauge group as $G$, and a possible flavor symmetry of the charged matter as $F$ (or $F_i$, when $F = F_1 \times \ldots \times F_n$). Charged matter will transform in representations $(\omega_i, \lambda_i)$ of these symmetries, which can be determined e.g. by analyzing in detail the singularities occurring in the compactification geometry \cite{Bershadsky:1996nh, Morrison:2011mb} or extracted from its genus 0 Gromov-Witten invariants \cite{Kashani-Poor:2019jyo}. With a few exceptions, the gauge group $G$ only occurs once in the Higgsing tree above a given base $\IF_n$. We can hence identify these theories via the nomenclature $G_n$.\footnote{Specifying the flavor group is necessary only in the case of $n=1,2$, $G=D_6$ and $n=1$, $G=A_5$, see appendix \ref{a:higgsing_trees}.}

    The unique tensor field $B$ of a rank 1 theory is sourced by instantons of the 6d gauge group via a coupling 
    \begin{equation} \label{eq:tensor_instanton_coupling}
        \int B \wedge \tr F \wedge F  \,.
    \end{equation}
    Quantization around an instanton solution in 6d leads to a 2d soliton in the spectrum of the theory: a non-critical string.
        
     \subsection{Worldsheet theory of the non-critical string} \label{ss:worldsheet_theory}
   
    The effective description of the non-critical string of 6d theories without charged matter is given by a non-linear sigma model on the worldsheet $\Sigma$ of the string (the trivial directions of the instanton background) with target space the one instanton moduli space $\cM_{G,1}$ (see e.g. \cite{Tong:2005un} for a review). The 6d coupling \eqref{eq:tensor_instanton_coupling} induces the coupling
    \begin{equation}
        \int_\Sigma B \,.
    \end{equation}
    on the worldsheet of the string.

    Factoring out the center of mass motion, the reduced instanton moduli space $\widetilde \cM_{G,1}$ is a hyperk\"ahler manifold of quaternionic dimension $h_G^\vee -1$. Both the gauge group $G$ and spacetime rotations $SO(4)$ act isometrically on $\tM_{G,1}$, the former by modifying the embedding of the instanton into the gauge group by a global factor, the latter by changing the orientation of the instanton in spacetime. Furthermore, the $G$ action is triholomorphic, i.e. commutes with the hyperk\"ahler structure. Writing the spacetime rotations $SO(4) \sim SU(2)^L \times SU(2)^R$, the first factor is triholomorphic, while the second rotates the three complex structures into each other.

    A non-linear sigma model with hyperk\"ahler target space permits $N=4$ supersymmetry \cite{Alvarez-Gaume:1981exv,Howe:1987qv,Howe:1988cj}, and indeed, a fermion in a one-instanton background exhibits $4h_G^\vee$ right-moving zero modes. The worldsheet theory of the non-critical string should therefore exhibit $(0,4)$ supersymmetry, implying that the string is a BPS object of the 6d theory which breaks half of its supersymmetry. The fact that $SU(2)^R$ rotates the complex structures of $\tM_{G,1}$ implies that it also acts as an R-symmetry on the corresponding supercharges. 

    To incorporate the effect of the presence of charged matter in the 6d theory, we need to introduce fields that transform both under the symmetry $G$ which acts on $\cM_{G,1}$ and the flavor symmetry $F$. A natural proposal \cite{DelZotto:2018tcj} for a minimal modification of the matterless case is to invoke fields in $(0,4)$ Fermi multiplets;\footnote{This is also how matter is introduced in theories for which a UV description is known.} this introduces left-moving fermions, which we take to transform in the appropriate $G$-equivariant vector bundle over $\cM_{G,1}$ determined by the representations $(\omega_i, \lambda_i)$ of the charged matter under the gauge and flavor symmetry of the 6d theory. 

    An elegant argument for obtaining the central charge of the theories along Higgsing trees for $n=3,4,5,6,8,12$ is presented in \cite{DelZotto:2018tcj}: for the matterless theories, the $4(h_G^\vee -1)$ bosons contribute $4(h_G^\vee -1)$ to both $c_L$ and $c_R$, while the right-moving fermions contribute $\frac{1}{2}4(h_G^\vee -1)$ to $c_R$. Hence, for theories without charged matter,
    \begin{equation}
        c_L = 4(h_G^\vee -1) \,, \quad c_R = 6 (h_G^\vee -1 ) \,.
    \end{equation}
    The right-moving central charge should retain this form also for theories with charged matter, as these differ only by a left-moving fermion bundle. To obtain the left-moving central charge of these theories, invoke the invariance of the gravitational anomaly $c_L-c_R$ under Higgsing to obtain
    \begin{equation}
        c_L = (c_L - c_R)(G_0) + c_R(G) \,.
    \end{equation}
    Here, $G_0$ denotes the gauge group of the theory without charged matter at the base of the Higgsing tree (this is where the constraint on $n$ enters). For such theories, $h_{G_0}^\vee = 3(n-2)$ \cite{Shimizu:2016lbw}. Hence,
    \begin{equation} \label{eq:central_charges}
        c_L = 6(h_G^\vee -n) +8 \,, \quad c_R = 6(h_G^\vee -1) \,.
    \end{equation}
    Reading off the number of left-moving bosons and fermions in the theory from this results allows us to infer the left-moving Casimir energy $E^0$ of these theories in the R- and the NS-sector. We follow the mnemonic presented e.g. in \cite{Polchinski:1998rq}: a contribution of $-\frac{1}{24}$ per boson, $\frac{1}{24}$ for a periodic fermion (R-sector), $-\frac{1}{48}$ for an anti-periodic fermion (NS-sector). This yields
    \begin{equation} \label{eq:casimir_energies}
        \casimirR = \frac{7}{6} - \frac{n}{2} \,, \quad \casimirNS = -\frac{c_L}{24} \,.
    \end{equation}
    The result \eqref{eq:central_charges} can alternatively be derived using anomaly arguments, as we review in section \ref{ss:anomaly}.
   
    \subsection{Symmetries of the string worldsheet} \label{ss:symmetries_of_string_worldsheet}

    As is familiar from the critical string, both the gauge symmetry $G$ and the flavor symmetry $F$ of the 6d spacetime theory manifest themselves as global symmetries of the worldsheet. As we argued in section \ref{ss:worldsheet_theory} above, the worldsheet theory also inherits half of the $(1,0)$ supersymmetry of the ambient theory. The latter can be decomposed with regard to the directions tangent and normal to the string worldsheet; a (1,0) spinor thus decomposes as
	\begin{equation} \label{eq:decomp_10}
	    (1,0) \rightarrow (\boldsymbol{2},\boldsymbol{1},+) \oplus (\boldsymbol{1},\boldsymbol{2},-) \,,
	\end{equation}
	with regard to $SO(4) \times SO(1,1)= \frac{SU(2)_L \times SU(2)_R}{\IZ_2} \times SO(1,1)$, where we have indicated the two spinor chiralities of $SO(1,1)$ with $\pm$. 

    Perhaps somewhat surprisingly for minimal supersymmetry, the $(1,0)$ supersymmetry algebra exhibits an R-symmetry, i.e. a linear action on its supercharges. One way to describe this action is by first noting that \cite{Kugo:1982bn}
	\begin{equation}
	    SL(2, \IH) \xrightarrow[]{2:1} SO(1,5) \,.
	\end{equation}
	The spin representation in 6d can thus be realized as a rank 2 module $S_6$ over the quaternion algebra $\IH$. The group $SL(1,\IH) \cong SU(2)$ acts on $S_6$ via multiplication on the right. This action gives rise to the R-symmetry which we shall denote as $SU(2)_I$. In terms of it, a (1,0) spinor transforms in the $\boldsymbol{2}$ representation:
	\begin{equation}
	    S_6 \ni \psi  = \begin{pmatrix} \mu \\ \nu \end{pmatrix} \,, \quad \mu, \nu \in \IH \,,
	\end{equation}
	with $\Lambda \in SL(2,\IH)$ and $M \in SL(1,\IH)$ acting as 
	\begin{equation}
	    \psi \mapsto \Lambda \cdot \psi \cdot M \,.
	\end{equation}
	To map to modules over $\IC$, we realize the quaternionic generators $i,j,k$ in terms of the matrices $i \sigma_1, i \sigma_2, i \sigma_3$. $\psi$ is then represented by a $2 \times 4$ matrix of complex numbers. Each of the four rows of this matrix transforms in the $\boldsymbol{2}$ of $SU(2)$. The diagonal action on all four rows commutes with the action of the Lorentz group.
	
	Returning to the decomposition \eqref{eq:decomp_10}, the representation under the R-symmetry $SU(2)_I$ goes along for the ride, yielding \cite{Shimizu:2016lbw}
	\begin{equation}
    	 (1,0,\boldsymbol{2}) \rightarrow (\boldsymbol{2},\boldsymbol{1},+,\boldsymbol{2}) \oplus (\boldsymbol{1},\boldsymbol{2},-,\boldsymbol{2}) \,.
	\end{equation}
	The string breaks half of the ambient supersymmetry. Taking the second summand to be the one that is conserved (this is a choice of instanton vs. anti-instanton), this gives rise to $(0,4)$ supersymmetry on the worldsheet, with the $SO(4)$ R-symmetry identified with $SU(2)_R \times SU(2)_I/\IZ_2$.\footnote{The $\IZ_2$ quotient reflects the fact that the negative of the diagonally embedded identity into $SU(2)_R \times SU(2)_I$ acts trivially on the spinor.} 
    
    In section \ref{ss:worldsheet_theory}, we identified a symmetry $SU(2)^L \times SU(2)^R$ of the reduced one-instanton moduli space $\widetilde \cM_{G,1}$. $SU(2)^L$ is naturally identified with $SU(2)_L$. $SU(2)^R$ however acts both as a right chiral spacetime rotation and an R-symmetry. It is hence naturally identified with a diagonal embedding into $SU(2)_R \times SU(2)_I$. This embedding will play a role in identifying the fugacities of the elliptic genus of the worldsheet theory with couplings of the topological string in section \ref{ss:fugacities} below.

    \subsection{What anomaly inflow teaches us about the worldsheet theory} \label{ss:anomaly}
    
    6D $\mathcal N=(1,0)$ theories are generically anomalous with an anomaly polynomial given by \cite{Shimizu:2016lbw,Sadov:1996zm,Sagnotti:1992qw}
    \begin{equation*}
        \frac{1}{2}\eta^{ij}I_i\wedge I_j\,,
    \end{equation*}
    where $\eta^{ij}$ is the charge pairing and
    \begin{equation*}
        \eta^{ij}I_j=\frac{1}{4}\left(\eta^{ia}\tr F_a^2-(2-\eta^{ii})p_1(T)\right)+h_{G_i}^\vee c_2(I)\,,
    \end{equation*}
    with $a$ indexing both the dynamical and the background fields, $c_2(I)$ the Chern class of the $SU(2)_I$ bundle and $p_1(T)$ the first Pontrjagin class of (the tangent bundle of) $M^{(6)}$ \cite{Shimizu:2016lbw}. This anomaly can be canceled via the addition of a Green-Schwarz term in the action,
    \begin{equation}
        S_{GS}=\int_{M^{(6)}}\eta^{ij} B_i\wedge I_j \,,
        \label{eq:GS_action}
    \end{equation}
    and a modification of the conservation law for $B_i$: the invariant field is not $dB_i$, but $H_i=dB_i+\alpha
    _i$ with $d\alpha_i=I_i$.
    
    This extra term needed to cancel the anomalies gives a contribution to the 2D theory of the non-critical string. To get its 2D contribution, one considers the 6D theory in the presence of the string. As the string sources the $B$-field, the Bianchi identity of $H$ gets modified to
    \begin{equation}
        dH_i=I_1+Q_i\prod_{j=2}^5\delta(x_j)dx^j\,,
        \label{eq:background}
    \end{equation}
     where $Q_i$ is the charge of the string, and we have put the string in the $x_0,\,x_1$ plane. In the background given by this $H$ field, the Green-Schwarz term \eqref{eq:GS_action} combines with the kinetic term of the $B_i$ fields to give a contribution to the 2D worldsheet action of the string. This contribution can be computed using anomaly inflow \cite{Sadov:1996zm,Shimizu:2016lbw} and, of course, is not gauge/diffeomorphism invariant. 
    
    The contribution of this Green-Schwarz term to the 2D anomaly polynomial is given by 
    \begin{equation}
        I_{4}=\frac{\eta^{ij}Q_iQ_j}{2}(c_2(L)-c_2(R))+\eta^{ij}Q_iI_j\,,
    \end{equation}
    where $c_2(L/R)$ are the Chern classes of the $SU(2)_{L/R}$ bundles introduced above. Of course in this last equation one needs to decompose the 6D characteristic classes appearing in $I_i$ in terms of their 2D counterparts:
    \begin{equation*}
        p^{6D}_1(T)=p_1^{(2D)}(T)+p_1^{(4D)}(N)=p_1^{(2D)}(T)-2 c_2(L)-2c_2(R)\,.
    \end{equation*}
    We then arrive at 
    \begin{equation}
        I_{4}=\frac{\eta^{ij}Q_iQ_j}{2}(c_2(L)-c_2(R))+Q_i\left(\frac{1}{4}\eta^{ia}\tr F_a^2-\frac{2-\eta^{ii}}{4}(p_1(T)-2c_2(L)-2c_2(R))+h^\vee_{G_i}c_2(I)\right)\,.
    \end{equation}
    
    Rank 1 theories exhibit a single tensor field $B$. In terms of the F-theory engineering geometry, the pairing $\eta^{11}=n$ is given by the negative self-intersection number of the only compact curve in the base of the elliptic fibration. Focusing on the elliptic genus of a single string, we set $Q=1$ to obtain
    \begin{align}
        I_4&=-\frac{2-n}{4}p_1(t)+\frac{n}{4}\tr F_G^2+\frac{1}{4}\eta^{1a}\tr F_a^2+(c_2(L)+c_2(R))+(h_G^\vee c_2(I)-n c_2(R))\,.
        \label{eq:AnomalyPol}
    \end{align}
    This equation encodes much non-trivial information regarding the 2d theory \cite{DelZotto:2018tcj}, as we now review. This is the anomaly polynomial of the non-linear sigma model on the moduli space of instantons $\mathcal M_{G,1}$. As the center of mass multiplet gives a universal contribution, it is convenient to factor it out and discuss the reduced theory on the reduced moduli space $\tM_{G,1}$. We use a superscript $^\cM$ when referring to quantities associated to the full theory.
\begin{itemize}
    \item \textbf{Central charges:}
    
    The difference $c_L^\cM-c^\cM_R$ of the central charges is fixed by the gravitational anomaly to be the coefficient of $-\frac{1}{24}p_1(T)$:
    \begin{equation} \label{eq:difference_c}
        c_L^\cM-c_R^\cM=6(2-n)\,.
    \end{equation}
    Moreover, the central charge on the right is linked to the level of the R-symmetry in the IR. This is identified with $SU(2)_I$. Its level is given by the coefficient of $c_2(I)$ in the anomaly polynomial: $k_R=h_G^\vee$. This fixes the right-moving central charge to $c_R=6k_R$ \cite{Putrov:2015jpa,Schwimmer:1987,Witten:1997yu}, allowing us to also extract the left-moving central charge from \eqref{eq:difference_c}, yielding
    \begin{equation}
        c_L^\cM=6( h_G^\vee-n+2)\,,\quad c_R^\cM=6h_G^\vee\,.
    \end{equation}
    Subtracting the $c=(4,6)$ contribution from the center of mass motion, we arrive at
    \begin{equation}
        c_L=6( h_G^\vee-n+2)-4\,,\quad c_R=6h_G^\vee-6\,.
    \end{equation}
    \item \textbf{Current levels:}
    
    From the WZW-models (for a review see \cite{francesco2012conformal}), we know that the gauge part of the anomaly polynomial is proportional to the level of the gauge current. We thus conclude that the gauge algebra current is at level $-n$ while the flavor algebra currents are at level $-\eta^{1a}\geq 0$.
    The authors of \cite{DelZotto:2018tcj} determine the flavor algebras and levels for most of the theories in the class we consider. These results are reproduced in appendix \ref{sec:HiggsingTrees}.
    \begin{equation*}
        \text{Gauge algebra level }=-n\,,\quad\text{flavor algebra level }=-\eta^{ia}\geq 0\,.
    \end{equation*}
    \item \textbf{$SU(2)^L$ and $SU(2)^R$ levels:}
    
    As we argue in section \ref{ss:symmetries_of_string_worldsheet}, the $SU(2)_L$ and $SU(2)^L$ symmetries coincide. We can hence identify the coefficient of $c_2(L)$ in the anomaly polynomial with the level of the $SU(2)^L$ current. As this coefficient is 1, we conclude that all dependence on the associated fugacity $x$ is captured by the center of mass contribution to the elliptic genus, i.e.
    \begin{center}
        the fugacity $x$ does not appear in $\E$\,.\footnote{Note that this is no longer true for the elliptic genus of multiple strings.}
    \end{center}
    On the other hand, we identified the $SU(2)^R$ symmetry with the diagonal of $SU(2)_R\times SU(2)_I$. The level $\kappa^\cM$ of the corresponding current is therefore given by the sum of the coefficients of $c_2(I)$ and $c_2(R)$. This gives
    \begin{equation}
        \kappa^\cM=h^\vee_G-n+1\,,
    \end{equation}
    Removing the contribution of the center of mass multiplet yields
    \begin{equation}
        \kappa=h_G^\vee-n\,.
        \label{eq:kappa_def}
    \end{equation}

\end{itemize}

    \subsection{Fugacities of the elliptic genus and relation to topological string} \label{ss:fugacities}

    As reviewed in the appendix, the $(0,2)$ elliptic genus permits the inclusion of charges in the trace which commute with the two supercharges; the resulting index is sometimes referred to as a flavored elliptic genus. Natural choices are the exponentiated Cartan generators of the gauge and flavor symmetries $G$ and $F$, as well as of the chiral spacetime rotation $SU(2)^L$. We will call the associated fugacities as $m_G$, $m_F$, and $\epsilon_{-}$. For the exponentiated fugacities, we use $x=e^{2\pi i \epsilon_-}$ for the $SU(2)^L$ group and $Q_i=e^{2\pi i(\alpha_i,m)}$ or $X_i=e^{2\pi i (e_i,m)}$ for $G$ and $F$. Here, $\alpha_i$ are the simple roots and $e_i$ are the vectors providing the canonical basis of the associated Euclidean space. The conventions for the embedding of the coroot lattices of Lie groups in Euclidean lattices are the same as in \cite{Duan:2020imo}.
    
    As the worldsheet exhibits $(0,4)$ supersymmetry, an inclusion of generators for a subgroup of the R-symmetry with which two supercharges commute is also permissible. Over the reals, the best we can do is consider an embedding
    \begin{equation}
        U(1) \times U(1) \hookrightarrow SU(2)_R \times SU(2)_I \,.
    \end{equation}
    The insertion of the generator of one of the $U(1)$ factors into the trace determines which subset of 2d BPS states contributes to the elliptic genus. Let $J^3_R$ and $J^3_I$ indicating the infinitesimal generators of the Cartan of $SU(2)_R$, $SU(2)_I$ respectively. To connect to partition functions in the $\Omega$ background \cite{Nekrasov:2002qd} and the refined topological string partition function \cite{Iqbal:2007ii,CKK}, we would like to introduce a fugacity $v$ associated to the symmetry $SU(2)^R$ with Cartan $J^3_R + J^3_I$ by inserting
    \begin{equation}
        X_v = v^{J^3_R + J^3_I}
    \end{equation}
    into the elliptic genus. Without this insertion, only states whose right-moving factor is annihilated by all supercharges contribute. Upon insertion, only annihilation by supercharges which commute with $X_v$ suffices.

    \section{Three constraints on the elliptic genus and an intermezzo} \label{s:four_constraints}

    \subsection{The modular ansatz}  \label{ss:modular_ansatz}
    The (flavored) elliptic genus is essentially a meromorphic Jacobi form of vanishing weight and of index fixed by the 't Hooft anomalies governing the flavor symmetries \cite{Benini:2013a, Benini:2013xpa}. We say essentially, because in theories with a gravitational anomaly, i.e. for which left and right moving central charge do not coincide, the defining transformation properties for Jacobi forms under modular transformations are modified by a phase, see e.g. equation (2.16) in \cite{Benini:2013xpa}.
    
    For the theories under consideration, the R-R elliptic genus has been argued \cite{Huang:2015sta,  DelZotto:2016pvm, Gu:2017ccq, DelZotto:2017mee, DelZotto:2018tcj} to take the form 
    \begin{equation}
        \ER=\eta(q)^{24 \casimirR} \frac{N(q,v,m_G,m_F)}{D(q,v,m_G)}\,,
        \label{eq:modular_all_fugacities}
    \end{equation}
    with $N$ and $D$ polynomials in Weyl invariant holomorphic Jacobi forms \cite{Wirthmuller:Jacobi,Bertola},\footnote{We are considering weak Jacobi forms to be a special case of Weyl invariant holomorphic Jacobi forms, see e.g. the appendix of \cite{DelZotto:2017mee}.} and $\casimirR$ the left-moving Casimir energy in the Ramond sector.\footnote{\label{fn:shift}For theories over the Hirzebruch base $\IF_1$, the power of the $\eta$ function is actually the Casimir energy minus one. All statements in the rest of this paper remain true for the $n=1$ case if we interpret $\casimirR$ as the Casimir energy minus one in this case, and we will henceforth tacitly do so.} The denominator $D$ in \eqref{eq:modular_all_fugacities} for a given group $G$ is universal. The numerator $N$ is a holomorphic Jacobi form of weight equal the weight of $D$ minus $12 \casimirR$ (the contribution of the Dedekind $\eta$ function to the weight), and of index in the elliptic parameters $v$ and $G$ adjusted by the index of $D$. As the space of Weyl invariant holomorphic Jacobi forms of fixed weight and indices is a finite dimensional vector space,\footnote{This is true with the exception of the space of Weyl invariant Jacobi forms for the Weyl group of $E_8$ \cite{Wirthmuller:Jacobi}, see \cite{Sakai:2011xg,Sakai:2017ihc, Wang:2018fil,Sun:2021ije} for this case.} $N$ is fixed once a finite number of coefficients are determined. Unfortunately, the number of coefficients is typically very large: there are e.g. 236,509 terms for the $(F_4)_4$ theory. This number is in many cases considerably reduced by invoking multiple sources of symmetry enhancement \cite{Duan:2020imo} to express $N$ in terms of Jacobi forms of Weyl groups larger than that of $G$.

    The coefficients required to determine $N$ were fixed in \cite{Huang:2015sta, Gu:2017ccq, DelZotto:2017mee} by matching to boundary conditions provided by the topological string partition function. \cite{Kim:2018gak} determined them by matching to the elliptic genus expressed in terms of $\theta$-functions for theories for which a UV description is known. \cite{DelZotto:2018tcj} instead matched to universal features of the elliptic genus inspired by the IR theory (mainly for vanishing gauge and flavor fugacities). This latter approach is naturally integrated into our computation of the elliptic genus in the affine expansion \eqref{eq:affineAnsatzIntro}, and we shall review it below.

    Note that the presentation of $\ER$ in terms of Jacobi forms as in \eqref{eq:modular_all_fugacities} is exact, just like the $\theta$-function expressions arising from knowledge of a UV theory. To compare to the topological string partition function, this expression must be expanded for small $q$, and then for small exponentiated gauge and flavor fugacities. This leads (though non-trivially) to an expansion in only positive exponentiated K\"ahler classes, as behooves the topological string partition function \cite{Gu:2017ccq}. This expansion also preserves the $v \rightarrow \frac{1}{v}$ symmetry of \eqref{eq:modular_all_fugacities}. In contrast, the presentation in terms of affine characters that we are interested in requires the expansion in small $q$ to be followed up by an expansion in small $v$. The contribution to the elliptic genus that is thus sensitive to the expansion region is that stemming from the zero modes. Already in \cite{Witten:1993jg}, these were seen to require a careful treatment. It would be desirable to understand the relation between expansion region and physical interpretation of the elliptic genus better. 

    \subsection{The affine ansatz} \label{ss:affine_ansatz}
    The central contribution of this note is providing ample evidence for the conjecture that the R-R elliptic genus for the class of theories described in section \ref{s:theories_considered} can be parametrized as follows:
    \begin{equation} \label{eq:flavor_sum}
        \E = \sum_{\lambda \in \dominantflavor} \hat{\chi}^F_\lambda(m_F,q) \, \xi^{n,G}_\lambda(m_G,q,v)
    \end{equation}
    with
    \begin{equation} \label{eq:affine_G_expansion}
    \xi_{\lambda}^{n,G}=\frac{1}{\eta(q)}\sum_{\omega \in \dominantgauge} \hat\chi_\omega(m_G,q) \sum_{k\in \mathbb Z} c^\lambda_{\omega,k} \,  q^{-\frac{k^2}{4\kappa}}v^k \,,
    \end{equation}
    with $c^\lambda_{\omega,k}\in \{0,\pm 1,\pm 2\}$ and non-vanishing at fixed $\omega$ and $\lambda$ for only finitely many integers $k$. This ansatz is heavily inspired by the work \cite{DelZotto:2018tcj}. Several comments are in order. 
    \begin{itemize}
        \item The sum in equation \eqref{eq:flavor_sum} is over all highest weight representations $L_\lambda$, $\lambda$ dominant, of the affine Lie algebra associated to $F$ at a fixed positive level $k_F$. As the level $k$ of a rank $r$ affine Lie algebra is related to the Dynkin labels $\lambda_i$ of its weights via
        \begin{equation} \label{eq:levelDynkin}
            k = \sum_{i=0}^r a_i^\vee \lambda_i \,,
        \end{equation}
        with the $a_i^\vee$ denoting the (non-negative) co-marks of the Lie algebra, this sum is necessarily finite. The notation $\hat \chi_{\lambda}^F$ indicates the affine character associated to the representation $\lambda$, normalized such that the leading power in $q$ is 
        \begin{equation}
            -\frac{c_F}{24}+h_\lambda^F \,,
        \end{equation}        
        with $c_F$ the central charge of the WZW model associated to the affine Lie algebra at this level, and $h_\lambda^F$ the conformal weight of a state with highest weight $\lambda$ in this theory. The explicit expressions are
        \begin{equation}
            c_F=\frac{k_F \dim(F)}{h^\vee_F+k_F}\,, \quad h_\lambda^F=\frac{\left<\lambda,\lambda+2\rho_F\right>}{2(h_F^\vee+k_F)}\,,
        \end{equation}
        where the inner product $\langle \cdot, \cdot \rangle$ is normalized such that long roots have length $2$.
        
        \item The sum over representations $\omega$ in equation \eqref{eq:affine_G_expansion} is over highest weight representations $L_\omega$ of the affine Lie algebra $G$ at negative level $-n$, constrained as follows: by the relation \eqref{eq:levelDynkin}, $\omega$ cannot be dominant. We require its finite projection to be dominant. This permits us to impose that the finite representations to which $L_\omega$ restrict at each grade be integrable, thus salvaging at least the symmetry under the finite part of the Weyl group.
        
        As $\lambda_0$ can become arbitrarily negative, the sum over $\omega$ is infinite. The affine characters $\hat \chi_\omega$ are normalized such that the leading power in $q$ is 
        \begin{equation}
            -\frac{c_G}{24}+h_\omega^G
        \end{equation}         
        with 
        \begin{equation} \label{eq:c_and_h_for_omega}
            c_G=\frac{\dim(G)(-n)}{\kappa}\,, \quad h_\omega^G=\frac{\left<\omega,\omega+2\rho_G\right>}{2\kappa}\,,
        \end{equation}
        and $\kappa$ given in equation \eqref{eq:kappa_def}. We discuss these characters and their computation further in appendix \ref{a:Kazhdan-Lusztig}.

        \item We will argue below that for fixed $\omega$, the sum over $k$ is finite. $\xi_{\lambda}^{n,G}$ thus has the structure
        \begin{equation} \label{eq:xi_in_terms_of_p}
        \xi_{\lambda}^{n,G}=\frac{1}{\eta(q)}\sum_{\omega \in \dominantgauge} \hat\chi_\omega(m_G,q) \,p_\omega^\lambda(q^{h_v^1},v) \,,
        \end{equation}
        where we have defined
        \begin{equation} 
            h_v^k = -\frac{k^2}{4\kappa} 
        \end{equation}
        and $p_\omega^\lambda(y,v)$ denotes a polynomial of the form
        \begin{equation} \label{eq:poly_multiplying_omega}
            p_\omega^\lambda(y,v) = \sum_k c^\lambda_{\omega,k} \,v^k y^{k^2} \,.
        \end{equation}
        \end{itemize}
    The computational task required to obtain $E_R$ is to determine which representations $\omega$ contribute, and for each such $\omega$, to compute the polynomials $p_\omega^\lambda(y,v)$.

    \subsection{Intermezzo: from $\ER$ to $\ENS$} \label{ss:intermezzo}
    Both for deriving a lower bound on the negative powers of $v$ that occur in $\ER$, and to obtain some constraints on the expansion coefficients, it will prove useful to be able to relate $\ER$ to the NS-NS elliptic genus. Surprisingly (recall that we have fermions in the left-moving, non-supersymmetric sector), this relation is (conjecturally) delightfully simple: the following relation is conjectured in \cite{DelZotto:2016pvm,DelZotto:2018tcj}:\footnote{The authors of \cite{DelZotto:2018tcj,DelZotto:2016pvm} conjecture this map to yield the elliptic genus in the NS-R sector. This is not consistent with level matching, due to the ensuing occurrence of both integral and half-integral powers of $q$ relative to the Casimir energy. This structure is consistent with the NS-NS elliptic genus, due to the shift \eqref{eq:anticommutation_NS} by the right-moving R-current  $\bar J_0$ in the power of  $\bar q$ in that case. We thank Michele Del Zotto for a discussion regarding this point.}
    \begin{equation}
        \ENS (m_G,m_F,v,q)=q^{-\kappa/4}v^\kappa \E(m_G,m_F,q^{1/2}/v,q), \quad \kappa=h_G^\vee - n. \label{eq:NSR-R relation}
    \end{equation}
    Note that this is a priori not a manifestation of spectral flow of $N=2$ theories, as the left-hand side of the theories we are considering do not exhibit supersymmetry. However, in a theory of bosons and fermions with a $U(1)$ charge with generator $J_0$, one may try to relate traces over R and NS sectors by shifting $L_0$ with a multiple of the generator $J_0$ to account for the half-integral moding of fermions in the NS sector compared to the R sector, and shifting both $L_0$ and $J_0$ to account for the different weight and charge of the vacuum in the NS vs. the R sector. For the $N=2$ theory of free fermions and bosons, this procedure indeed reproduces spectral flow. Surprisingly, the same strategy reproduces \eqref{eq:NSR-R relation}. The shifts that lead to \eqref{eq:NSR-R relation} are
    \begin{equation} \label{eq:shifts_of_weight_and_charge}
        h_{\mathrm{NS}}=h_\mathrm{R}+\frac{1}{2}l_\mathrm{R}+\frac{1}{2}k\,, \quad l_{\mathrm{NS}}=l_\mathrm{R}+2 k \,,
    \end{equation}
    with $k= -\kappa/2$. Indeed, composing this shift with the $v \to 1/v$ symmetry of the elliptic genus in the form \eqref{eq:modular_all_fugacities}, we have 
    \begin{equation} \label{eq:RtoNS_monomial}
        q^{h_\mathrm{R}}v^{l_\mathrm{R}}\mapsto q^{h_\mathrm{R} + \frac{l_\mathrm{R}}{2} + \frac{k}{2}} v^{-l_\mathrm{R} -2k} =q^{k/2}v^{-2k}\left(\left.q^{h_\mathrm{R}} v^{l_\mathrm{R}}\right|_{v\mapsto q^{1/2}/{v}}\right) \,,
    \end{equation}
    which induces the transformation \eqref{eq:NSR-R relation}.

    It will be convenient to introduce the following notation: we will have $\mathcal F_\kappa[f]$ denote the function $f$ upon acting with the transformation \eqref{eq:RtoNS_monomial}, i.e. 
    \begin{equation} \label{eq:def_F}
        \mathcal F_\kappa [f](m_G,m_F,v,q):=q^{-\kappa/4}v^\kappa f(m_G,m_F,q^{1/2}/v,q) \,,
    \end{equation}
    such that 
    \begin{equation}
        \ENS = \mathcal F_\kappa[\E] \,.
    \end{equation}
    
    We will see below that $\E$ has an expansion in positive powers of $q/v^2$ and $v$. Disregarding the prefactor, the transformation \eqref{eq:def_F} on a monomial of this expansion is
    \begin{equation}
        \left(\frac{q}{v^2}\right)^j v^l \mapsto v^{2j} \left(\frac{q}{v^2}\right)^{l/2} \,.
    \end{equation}
    The expansion region, small $q^2/v$, small $v$, is hence preserved by $\mathcal F_{\kappa}$. To obtain a transformation with this property, we had to compose the shifts \eqref{eq:shifts_of_weight_and_charge} with the symmetry $v \rightarrow 1/v$ of the unexpanded elliptic genus.
    
The fact that $\ENS=  \mathcal F_\kappa(\E)$ implies that the same ansatz we had for $\ER$ holds for $\ENS$:\footnote{Note that $q^{-\kappa/4}v^\kappa \left.\left(v^\ell q^{-\frac{\ell^2}{4\kappa}}\right)\right|_{v\to q^{1/2}/v}= q^{-\frac{(\kappa-\ell)^2}{4\kappa}}v^{\kappa-\ell}$}
\begin{equation}
    \ENS = \sum_{\lambda \in \dominantflavor} \hat{\chi}^F_\lambda \, \xi^{\mathrm{NS}}_\lambda=
      \frac{1}{\eta} \sum_{\lambda \in \dominantflavor} \hat{\chi}^F_\lambda \, \sum_{\omega \in \dominantgauge} c^{\mathrm{NS},\lambda}_{\omega,z}\hat \chi_\omega(m_G,q) \sum_{\ell\in \mathbb Z} q^{-\frac{\ell^2}{4\kappa}}v^\ell.
      \label{eq:affineAnsatzNS}
\end{equation}

    For the cases without massless matter, the NS-NS and the R-R elliptic genus coincide (up to an irrelevant sign choice). Hence, the $\xi$ functions must be $\mathcal{F}_\kappa$ invariant. For the general case\footnote{$(E_7)_7$, the only theory with massless matter but no flavor group, is an exception to this rule, see subsection \ref{subsec:E7_7}. }, we will find that $\mathcal F_\kappa$ permutes the different $\xi$ functions, i.e.
\begin{equation}
    \mathcal F_\kappa (\xi^G_\lambda)=\pm\xi^G_{\lambda'} \quad \text{ for some } \lambda' \,.
    \label{eq:Fpermutation}
\end{equation}

    \subsection{The low lying spectrum in the NS sector} \label{ss:low_lying_spectrum}
    \cite{DelZotto:2016pvm} argue, building on previous observations in \cite{Benvenuti:2010pq} and \cite{Keller:2011ek,Keller:2012da,Hanany:2012dm,Rodriguez-Gomez:2013dpa}, that the leading contribution at $q^{-c_L/24}$ to the NS elliptic genus should essentially coincide with the Hilbert series of the corresponding one-instanton moduli spaces. \cite{DelZotto:2018tcj} observe that the contribution (in the presence of charged matter in the spacetime theory) at the next level, $q^{-c_L/24+1/2}$, also has a universal form in terms of (finite) characters of the groups $G$ and $F$.\footnote{This is observed for the examples $(C_r)_1$ for $r=1,2,3,4$, $(G_2)_3$, $(D_r)_4$ for $r=4,5$, and $(B_4)_4$.} We conjecture that this form is universally valid for the theories that we can consider and impose the ansatz
    \begin{equation} \label{eq:ansatz_low_lying}
        \ENS=q^{-\frac{c_L}{24}} v^{h_G^\vee-1}\left(\sum_{k}v^{2k}\chi^G_{k\theta}(m_G)-q^{\frac{1}{2}}\sum_{k}v^{2k+1}\sum_{i}\chi^G_{k\theta+\omega_i}(m_G)\chi^{F}_{\lambda_i}(m_F)+O(q)\right)
    \end{equation}
    as a constraint on the elliptic genera. Here, $\theta$ is the highest root of $G$, and the pairs ($\omega_i$, $\lambda_i$) of representations of $G$ and $F$ were discussed in section \ref{ss:spacetime_theory}. Let us comment on these two contributions: considering the NS-NS rather than the R-R elliptic genus disentangles the fermionic and bosonic contributions at the ground state energy $\casimirNS = -c_L/24$ calculated in \eqref{eq:casimir_energies}. The leading contribution thus arises from symmetric tensor products of bosonic zero modes. These carry the adjoint representation of $G$. The $k$-th such product gives rise to the contribution $\chi_{k\theta}^G(m_G)$. As
    \begin{equation}
        \left(\theta^{\otimes k} \right)_{\mathrm{Sym}} = k \theta + \ldots \,,
    \end{equation}
    the form \eqref{eq:ansatz_low_lying} encodes that all additional contributions in the decomposition of the tensor product do not contribute independently \cite{Benvenuti:2010pq}.\footnote{We thank Noppadol Mekareeya for explaining to us how all representations beside $k\theta$ are set to zero by $D$-term constraints in the case $G=A_n$.} We can further read off the $U(1)_v$ charge of the NS vacuum to be $h_G^\vee -1$, and the $U(1)_v$ charge of the bosonic fields of the non-linear sigma model to be 2. The contributions at energy $-c_L/24 + 1/2$ arise when $1/2$ modes of the left moving fermion fields act on the bosonic ground states. These carry the representations $(\omega_i, \lambda_i)$ under the groups $(G,F)$, and we can read off their $U(1)_v$ charge to be 1. We again see that only the leading term $k\theta \oplus \omega_i$ in the decomposition of $\theta^{\otimes k} \otimes \omega_i$ in irreducible representations contributes. 

\subsection{Imposing $|c^\lambda_{\omega,k}|\leq 2$}

We have already discussed the three main constraints on the elliptic genus: (1) the modular ansatz \eqref{eq:modular_all_fugacities}, (2) compatibility with the NS elliptic genus, and (3) the affine ansatz \eqref{eq:affineAnsatzIntro}. The modular ansatz has finitely many unknowns; hence, by providing a finite amount of initial data or vanishing conditions, it is possible, in principle, to solve for the elliptic genus \cite{Huang:2015sta,Gu:2017ccq,Duan:2020imo,DelZotto:2016pvm,DelZotto:2018tcj}. In this work, we bypass the computation of initial data (except for the data coming from \eqref{eq:ansatz_low_lying}), and conjecture that imposing the other constraints is enough to determine the entire elliptic genus. However, even for small rank groups, the dimension of the Jacobi ring to which the numerator $N$ belongs is very large, making the problem of solving for all coefficients computationally intractable. Therefore, we opt to specialize some of the elliptic parameters to 0, thus greatly reducing the dimension of the space of Jacobi forms. This process, of course, removes information, so we are not able to determine the complete set of coefficients $c^{\lambda}_{\omega,k}$ uniquely. We find experimentally, however, that imposing the constraint
\begin{equation}
|c^{\lambda}_{\omega,k}|\leq 2\
\label{eq:constraintC}
\end{equation}
allows us to do so.\footnote{Up to small subtleties. See section \ref{sec:algorithm}}

We believe that the constraint \eqref{eq:constraintC} holds generally, for the following reasons:
\begin{itemize}
\item It holds for the cases we can compute exactly: For a handful of theories, including $(C_r)_1$, $(B_r/D_r)_4$, and $(G_2)_3$, we know the full elliptic genus with all fugacities turned on \cite{Kim:2014dza,DelZotto:2018tcj,Haghighat:2014vxa,Kim:2018gjo,Kim:2016foj,Kim:2015fxa,Kim:2018gak}. For these theories, we did not impose \eqref{eq:constraintC}, but we observed that it was satisfied.
\item The constraint is compatible with Higgsing: As we will explain in section \ref{s:Higgs_tree_relations}, there is a simple rule to obtain the elliptic genus of a Higgsed theory from the elliptic genus of its parent. We consider the fact that Higgsing preserves this constraint as strong evidence for its validity.
\item Simplicity: The affine characters reflect the structure of a multiplet plus all of its descendants. The affine ansatz already captures the structure of the gauge and flavor groups, so the only remaining source for structure is the relatively simple supersymmetry algebra (only 4 generators). We therefore expect $c^\lambda_{\omega,k}$ to be relatively small numbers. For several cases, we slightly increased the bound (up to 8) and found no additional solutions. In some cases in which we did find solutions violating this bound, the bound was violated for some coefficients by many orders of magnitude.
\item Existence of a solution: Using the algorithm in section \ref{sec:algorithm}, we found a solution that satisfies the bound \ref{eq:constraintC} in every theory we studied. The equations satisfied by the $c$'s are linear equations with large coefficients (approximately $10^8$ in the larger cases). We consider the fact that solutions exist that satisfy the bound to be strong evidence that the constraint should hold in general.
\end{itemize}

    \section{Putting the constraints to work}    
    \label{sec:algorithm}
    \subsection{The strategy} \label{ss:strategy}
    In the expressions \eqref{eq:flavor_sum} and \eqref{eq:affine_G_expansion}, $E_R$ is obtained as a sum over irreducible highest weight representations $L_\lambda$ of the symmetry $F$ at positive level $k_F$, a sum of highest weight representations $L_\omega$ of the symmetry $G$ at negative level $-n$, and a sum over powers $k$ of the fugacity $v$. The sum over $\lambda$ is finite, as only finitely many irreducible integrable highest weight representations exist at given level. The sum over $\omega$ is infinite. We will now argue that at fixed $\lambda$ and $\omega$, the sum over $k$ is finite. In each summand, the powers of $q$ are manifestly integrally spaced, with the leading power $\fpq_0$ of $q$ in the summand indexed by $\lambda$, $\omega$, and $k$ given by
    \begin{equation}
        \lpq(\lambda,\omega,k):=-\frac{c_F}{24}+h_\lambda^F-\frac{c_G}{24}+h_\omega^G+h^v_{k}-\frac{1}{24} \,.
        \label{eq:qExponentEquation}
    \end{equation}
    Note that $h_k^v$ is a negative definite quadratic form of $k$. Thus, at given $\lambda$ and $\omega$, $\lpq$ can be bounded by the Casimir energy $\casimirR$ (computed in \eqref{eq:casimir_energies}) only for a finite set of $k$. This is why at given $\omega$, only a finite number of terms can occur in the sum over $k$ in the expression \eqref{eq:affine_G_expansion} for $\xi_\lambda^{n,G}$, i.e. $c_{\omega,k}^\lambda = 0$ for all but finitely many $k$. 
    
    The modular ansatz \eqref{eq:modular_all_fugacities} implies that all occurring powers of $q$ (i.e. not merely restricted to a given summand) be integrally spaced relative to the lower bound on this power provided by the Casimir energy $E_0^R$,\footnote{Recall our conventions regarding the notation $\casimirR$ for the case $n=1$ as stated in footnote \ref{fn:shift}.}
    \begin{equation}
        \lpq(\lambda, \omega, k) - E_0^R\,\in\, \mathbb N  \quad \forall \lambda, \omega, k \,.
    \label{eq:constraint1}
    \end{equation}
     This constraint also follows from level matching at the level of the worldsheet theory. It constrains the representations $\omega$ that can contribute to the expansion \eqref{eq:affine_G_expansion} at fixed order in $q$ and $v$. As $h_\omega^G$, given in equation \eqref{eq:c_and_h_for_omega} above, is a positive definite quadratic form of the Dynkin labels of $\omega$, the number of such representations compatible with the constraint \eqref{eq:constraint1} is finite. 

    Next, we will argue that the power of $v$ that occurs in the expansion of $\ER$ at a given order in $q$ is also bounded below. This argument relies on considering the elliptic genus in the NS-NS sector, and invoking the lower bound on the power of $q$ there. Recall that, as discussed in section \ref{ss:intermezzo}, the monomial $q^\fpq v^k$ contribution to $\ER$ is mapped to the contribution $q^{\fpq+k/2-\kappa/4}v^{\kappa-k}$ to $\ENS$. The NS-NS elliptic genus permits both integral and half-integral energy levels relative to the Casimir energy $\casimirNS=-\frac{c_L}{24}=-\frac{\kappa}{4}-\frac{1}{3}$,
    \begin{equation} \label{eq:bound_on_k}
        \fpq + \frac{k}{2} - \frac{\kappa}{4}-(\casimirNS)=(\fpq-\casimirR)+\frac{k}{2}-\frac{n-3}{2}\, \in\, \frac{1}{2}\mathbb N_0  \,,
    \end{equation}
    thus implying the sought after lower bound on the power of $v$. Introducing the non-negative integers $j$ and $\ell$ via  
    \begin{equation}
       \ell = k- (n-3) + 2(\fpq- \casimirR) = k- (n-3) + 2j
    \end{equation}
    allows us to write $\ER$ as
    \begin{equation}
       \E=q^{\casimirR}v^{n-3}\sum_{j,\ell \ge 0} b_{j,\ell}(m_G,m_F)\left(\frac{q}{v^2}\right)^jv^\ell \,.
     \label{eq:bexpansion}
    \end{equation}

    The relation \eqref{eq:bound_on_k} also implies a lower bound on the power $k$ of $v$ in the polynomials \eqref{eq:poly_multiplying_omega}:
    \begin{equation}
    k\geq (n-3) -2(\lpq(\lambda, \omega,k) - \casimirR)
        \label{eq:constraint2}
    \end{equation}
    recall that the $k$ dependence in $\lpq(\lambda, \omega, k)$ is via the summand $-\frac{k^2}{4\kappa}$.

    The discussion above clarifies how the expansion \eqref{eq:affine_G_expansion} is to be understood: at any fixed order in $q$ and in $v$, all occurring sums, in particular the sum of non-integrable representations $\omega$, are finite.

    Now that we understand the nature of the expansion, we can go about solving for the coefficients $c_{\omega,k}^\lambda$ in the ansatz \eqref{eq:affine_G_expansion} by equating the expansion \eqref{eq:flavor_sum} to the modular ansatz \eqref{eq:modular_all_fugacities}. To render the computation feasible, we set, following \cite{DelZotto:2018tcj}, the fugacities $m_G = m_F =0$. This greatly reduces the number of unknown coefficients in the numerator $N$ of the modular ansatz, as the Jacobi forms depending on these fugacities specialize to integers. The constraint on the coefficients $c_{\omega,k}^\lambda$ thus takes the form 
    \begin{equation}
        \eta^{24 \casimirR}\frac{N(q,v,m_G=0,m_F=0)}{D(q,v,m_G=0)}=\left.\sum_{\lambda \in \dominantflavor} \hat{\chi}^F_\lambda \, \xi^G_\lambda\right|_{m_F=m_G=0} \,,
    \label{eq:tosolve}
    \end{equation}
    where the LHS is expanded first in $q$ and then in $v$. This gives rise to a linear homogeneous equation on both the coefficients of the basis of Jacobi forms at appropriate weight and level contributing to $N$ as well as the expansion coefficients $c_{\omega,k}^\lambda$ occurring in $\xi_\lambda^G$. To introduce inhomogeneities, we impose the knowledge of the low lying spectrum in the NS sector as described in section \ref{ss:low_lying_spectrum}.

    In terms of the expansion given in equation \eqref{eq:bexpansion}, the low lying spectrum in equation \eqref{eq:ansatz_low_lying} fixes\footnote{For $n=1$, the conditions are slightly different due to the presence of the tachyon and the shift in $\casimirR$ we introduced in footnote \ref{fn:shift}. They read $b_{0,0}=1\,,b_{0,\ell}=0\,,b_{k,2}=\chi_{k\theta}^G\,,b_{k+1,3}=\sum_i\chi^G_{\omega_i+k\theta}\chi_{\lambda_i}^F$}
    \begin{equation}
        b_{k+n-2,0}=\chi^G_{k\theta}
    \quad \text{and}\quad b_{k+n-1,1}=\sum_{i}\chi^G_{\omega_i+k\theta}\chi^F_{\lambda_i}
        \label{eq:contraint3} \,;
    \end{equation}
    equivalently,
    \begin{align*}
        c^{\lambda}_{\omega,1-n-2k}&=\delta_{\lambda,0}\delta_{\omega,k\theta} &\text{ if } k-\left((n-3)-2(\lpq(\lambda,\omega,k)-\E^0)\right)=0,\\
        c^{\lambda}_{\omega,-n-2k}&=-\sum_i\delta_{\lambda_i,\lambda}\delta_{\omega,\omega_i+ k\theta} &\text{ if } k-\left((n-3)-2(\lpq(\lambda,\omega,k)-\E^0)\right)=1/2.
    \end{align*}
    Note that in all of the examples that we consider, we can solve for the coefficients of $N$ first, just by taking advantage of coefficients of monomials in $q$ and $v$ that we know to vanish on the RHS of equation \eqref{eq:tosolve}. In most cases, the remaining constraints have a unique solution (once Dynkin symmetry is addressed, see immediately below) upon imposing $|c^\lambda_{\omega,k}|\leq 2$.

    Finally, we wish to discuss a complication due to possible Dynkin symmetry when setting the gauge and flavor fugacities in equation \eqref{eq:tosolve} to zero : characters associated to weights related by Dynkin symmetry are equal in the $m\to 0$ limit. Explicitly, if we have a Dynkin symmetry $s\in \text{Dyn}(G)$ for any character $\hat \chi_{w}^G$, 
    \begin{equation*}
       \hat \chi^G_{s\omega}(m_G)=\hat \chi^G_{\omega}(s m_G) \Rightarrow  \hat \chi^G_{s\omega}(0)=\hat \chi^G_{\omega}(0).
    \end{equation*}
    Therefore, equation \eqref{eq:tosolve} can only be solved for the sum $c^\lambda_{\omega,k}+c^\lambda_{s\omega,k}$ rather than for the individual coefficients. In many cases, we expect the elliptic genus to be Dynkin symmetric, as this symmetry is inherited via Higgsing from a theory further up the Higgsing tree \cite{Duan:2020imo}.\footnote{The cases for which we do not expect Dynkin symmetry are $(D_6)_{3}\,,\,\,(D_5)_{3}$, the two theories $(D_6)_1$, and $(D_6)_{2}$ with flavor group $C_6\times(\mathrm{Ising})\times (\mathrm{Ising})$.} In such cases, we can set $c^\lambda_{\omega,k}=c^\lambda_{s \omega,k}$ to resolve the ambiguity. Dynkin symmetry with regard to the flavor group can also occur, as  for $s\in \text{Dyn}(F)$, we cannot differentiate between $c^\lambda_{\omega,k}$ and $c^{s\lambda}_{\omega,k}$ upon setting $m_F = 0$. In these cases, we can only solve for $\xi_\lambda+\xi_{s\lambda}$. In the case of $U(1)$ flavor groups, the $m_{U(1)}\to -m_{U(1)}$ leads to the same style of ambiguity as Dynkin symmetry. 

    We exemplify this latter ambiguity at the hand of the example $(E_6)_5$ in appendix \ref{aa:polynomialsE65}.
    
    \subsection{An example: $\Fff$}
    
    As an example, consider the theory $(F_4)_4$. The flavor group is $A_1$ at level $3$ \cite{DelZotto:2018tcj}. At this level, there exist 4 dominant highest weight representations. Their highest weights are
    \begin{equation}
        \lambda_i = (3-i) \,\Lambda_0^{A_1} + i\, \Lambda_1^{A_1} \,, \quad i = 0, \ldots, 3 \,,
    \end{equation}
    where we have denoted the fundamental weights of $\hat A_1$ by $\Lambda_0^{A_1}$ and $\Lambda_1^{A_1}$.

    We will consider the expansion of the $\xi$ functions in equation \eqref{eq:affine_G_expansion} to order 
    \begin{equation}
        O\left(q^{\casimirR}\left(\frac{q}{v^2}\right)^{M+1},v^{m+(4-3)+1}\right) \,, \quad M=3 \,,\,\,m=8 \,.
    \end{equation}
    From our discussion in section \ref{ss:strategy}, we know that at a given power in $q$, the powers of $v$ are bounded below by \eqref{eq:constraint2}. Furthermore, at each power of $q$ and $v$, the new representations that contribute to the $\xi$ functions, i.e. that did not already contribute at lower power of $q$, are given by solving \eqref{eq:constraint1}. We organize the calculation in terms of orders of $q$:

    \hspace{1cm}
    
    {\bf Leading order $q^{\casimirR}$}
    
    Specialized to the leading order $q^{\casimirR}$, \eqref{eq:constraint2} which bounds the power of $v$ from below reduces to $k\geq (4-3)=1$. Hence, the powers $k$ of $v$ that can arise at order $O\left(q^{\casimirR}\left(\frac{q}{v^2}\right)^1,v^{m+2}\right)$ are $-1\leq k \leq m=9$. To find the affine characters of $F_4$ that contribute at this order, we thus find the admissible weights $\omega \in \dominant(F_4)_{-4}$ which solve \eqref{eq:constraint1}, 
    \begin{equation}
        \lpq(\lambda_i,\omega,k) - \casimirR = 0 \,,
        \label{eq:0orderF_4_4}
    \end{equation}
    for these values of $k$ and each of the four flavor weights $\lambda_i$. The solutions are given in table \ref{tab:0orderF_4_4}.
    
    \hspace{0.5cm}

    {\bf Order $q^{\casimirR+1} $}
    
    At the next order in $q$, $q^{\E^0+1}$, equation \eqref{eq:constraint2} reads $k\geq (4-3)-2=-1$, so for every value of $k$ with $-1\leq k \leq (m+1)-2 = 7$\footnote{$(m+1)-2$, as $(q/v^2) v^{m+1}=qv^{m+1-2}$.} we invoke \eqref{eq:constraint1},
    \begin{equation}
        \lpq(\lambda_i,\omega,k) - \casimirR = 1 \,,
        \label{eq:1orderF_4_4}
    \end{equation}
    to identify the new  representations $\omega$ that contribute at this order: we find that there are no solutions (those identified in table \ref{tab:0orderF_4_4} of course contribute at all powers of $q$ above the ground state).  
    
\begin{table}[]
        \centering
        \begin{tabular}{c|ccccccccc}
 \diagbox[]{$\lambda$}{k} & 1 & 2 & 3 & 4 & 5 & 6 & 7 & 8 & 9 \\\hline
 $\lambda_0$ & - & - & - & - & - & - & \text{(0000)} & - & - \\
  $\lambda_1$ & - & - & - & - & - & - & - & - & - \\
  $\lambda_2$& - & - & - & - & - & - & - & - & \text{(0001)} \\
  $\lambda_3$& - & - & - & - & - & - & - & \text{(0000)} & - \\    
        \end{tabular}
        \caption{Solutions of equation \eqref{eq:0orderF_4_4}}
        \label{tab:0orderF_4_4}
    \end{table}

In general at order $q^{\E^0+l}$ the equations read:
\begin{equation}
    e^{4,F_4}_{\lambda_i}(\omega,k)-\E^0=l, \quad  (4-3)-2l\leq k \leq m+1-2l \,.
    \label{eq:lOrderF_4_4}
\end{equation}
The solutions for $\omega$ for these equations for $l=2,3$ are given in tables \ref{tab:2orderF_4_4} and \ref{tab:3orderF_4_4}.

\begin{table}[]
        \centering
        \begin{tabular}{c|ccccccccc}
 \diagbox[]{$\lambda$}{k} & -3 & -2 & -1 & 0 & 1 & 2 & 3 & 4 & 5 \\\hline
 $\lambda_0$ & \text{(0000)} & - & - & - & - & - & \text{(0000)} & - & - \\
  $\lambda_1$  & - & - & - & - & - & - & - & - & - \\
  $\lambda_2$ & - & - & - & - & - & - & - & - & - \\
 $\lambda_3$ & - & - & - & - & - & - & - & - & - \\
        \end{tabular}
        \caption{Solutions of equation \eqref{eq:lOrderF_4_4} for $l=2$} 
        \label{tab:2orderF_4_4}
    \end{table}        

\begin{table}[]
        \centering
        \begin{tabular}{c|ccccccccc}
 \diagbox[]{$\lambda$}{k} & -5 & -4 & -3 & -2 & -1 & 0 & 1 & 2 & 3 \\\hline
 $\lambda_0$ & \text{(1000)} & - & - & - & - & - & - & - & - \\
  $\lambda_1$ & - & \text{(0001)} & - & - & - & - & - & - & - \\
  $\lambda_2$ & - & - & - & - & - & - & - & - & - \\
  $\lambda_3$ & - & - & - & \text{(0000)} & - & - & - & \text{(0000)} & - \\
        \end{tabular}
        \caption{Solutions of equation \eqref{eq:lOrderF_4_4} for $l=3$} 
        \label{tab:3orderF_4_4}
    \end{table}

    At this order, the ansatz for the $\xi$ functions therefore reads
\begin{align}
\label{eq:F_4_4xi}
\eta\xi^{4,F_4}_{(0)}&=\hat{\chi }_{\text{(0000)}} \left(v^7 y^{49} c^{\text{(0)}}_{\text{(0000)},7}+\frac{y^9 c^{\text{(0)}}_{\text{(0000)},-3}}{v^3}+v^3 y^9
   c^{\text{(0)}}_{\text{(0000)},3}\right)+\frac{y^{25} \hat{\chi }_{\text{(1000)}} c^{\text{(0)}}_{\text{(1000)},-5}}{v^5}+\dots,\\
\eta\xi^{4,F_4}_{(1)}&=   \frac{y^{16} \hat{\chi }_{\text{(0001)}} c^{\text{(1)}}_{\text{(0001)},-4}}{v^4}+\dots,\\
\eta\xi^{4,F_4}_{(2)}&=   v^9 y^{81} \hat{\chi }_{\text{(0001)}} c^{\text{(2)}}_{\text{(0001)},9}+\dots,\\
\eta\xi^{4,F_4}_{(3)}&=  \hat{\chi }_{\text{(0000)}} \left(v^8 y^{64} c^{\text{(3)}}_{\text{(0000)},8}+\frac{y^4 c^{\text{(3)}}_{\text{(0000)},-2}}{v^2}+v^2 y^4 c^{\text{(3)}}_{\text{(0000)},2}\right)+\dots\,,
\label{eq:F_4_4xilast}
\end{align}
where $y=q^{-\frac{1}{4\kappa}}=q^{-1/20}$ as in equation \eqref{eq:poly_multiplying_omega}. To lighten the notation, we have indexed the affine characters and the coefficients $c_{\omega,k}^\lambda$ only with the finite Dynkin labels; the zeroth Dynkin label is then determined by the level.

From the expressions \eqref{eq:F_4_4xi}-\eqref{eq:F_4_4xilast} it is difficult to read off at a glance the power of $q$ at which each term contributes to the elliptic genus, as both the affine characters associated to the flavor and to the gauge group exhibit a non-trivial leading $q$-power. Consider e.g. $\xi_{(1)}^{4,F_4}$:
\begin{align}
    x_0(\lambda=(1),\omega=(0001),k=-4) &= -\frac{c_{A_1}}{24}+h_{\lambda}^{A_1}-\frac{c_{F_4}}{24}+h_{\omega}^{F_4}+h^v_{-4}-\frac{1}{24}\\
&= -\frac{9/5}{24}+\frac{3}{20}-\frac{-208/5}{24}+\frac{6}{5}-\frac{16}{20}-\frac{1}{24}\\
&=\frac{13}{6}\\
&=3+\casimirR
\end{align}
where we have used $\casimirR=-\frac{5}{6}$. The first two terms come from the flavor character, the following two come from the gauge character, the $h^v_{-4}$ comes from $y^{16}$, and the $-\frac{1}{24}$ is the contribution from the Dedekind $\eta$-function.

    Before imposing equality with the modular ansatz as in \eqref{eq:tosolve}, we can fix some of the unknown constants from the knowledge of the low energy spectrum \eqref{eq:ansatz_low_lying} in the NS sector.

For the ${(F_4)}_4$ example \eqref{eq:F_4_4xi} these conditions fix three coefficients:
\begin{equation*}
    c^{\text{(0)}}_{\text{(0000)},-3}=1,\quad c^{\text{(0)}}_{\text{(1000)},-5}=1,\quad c^{\text{(1)}}_{\text{(0001)},-4}=-1.
\end{equation*}

For the ${(F_4)}_4$, the modular numerator $N$ has 61 undetermined coefficients, i.e. the Jacobi ring at the desired weight and index has dimension 61. Using the fact that small/negative powers of $v$ are constrained by \eqref{eq:bexpansion} we can solve for 42 coefficients and using \eqref{eq:contraint3} we solve for an extra 12 coefficients. The remaining 7 coefficients can be fixed by comparing the modular and the affine ansatz. Once the modular ansatz is fixed, we find a series of equations for the coefficients $c^\lambda_{\omega,k}$. Imposing $c^\lambda_{\omega,k}\in \mathbb Z$ with $|c^\lambda_{\omega,k}|\leq 2$ gives a unique solution.

\section{Closed form results for $\ER$} \label{s:closed_form_results}
For a subset of the theories we consider, we have conjectural closed form results for the elliptic genus $\ER$  in the form \eqref{eq:flavor_sum}: we enumerate all of the representations $\omega$ that contribute, and the associated polynomials $p_\omega^\lambda$ as defined in equation \eqref{eq:poly_multiplying_omega}. We present these results in this section. We have computed the representations $\omega$ that contribute to $\xi^{n,G}_\lambda$ to a certain fixed order in $q$ and the associated polynomials $p_\omega^\lambda$ for a host of other examples $G_n$, but we are not confident that to the order achieved, our results reflect the complete structure of $\ER$. These results are presented in appendix \ref{a:polynomials}.

Note that explicit results for the elliptic genera of the theories $(C_r)_1$, $(B_r/D_r)_4$, $(G_2)_3$ among others are known in terms of $\theta$ functions or modular forms are known \cite{Kim:2014dza,DelZotto:2018tcj,Haghighat:2014vxa,Kim:2018gjo,Kim:2016foj,Kim:2015fxa,Kim:2018gak}. Expressing these results in terms of our affine ansatz \eqref{eq:affineAnsatzIntro} (for $(C_r)_1$ below, and $(C_3)_1$, $(C_4)_1$, $(B_4)_4$, and $(D_5)_4$ in the appendix) hence provides evidence for the validity of this ansatz.\footnote{In practice, we only check the result with completely arbitrary fugacities to low orders due to the computational complexity of the problem. To arrive at the results we give in appendix \ref{a:polynomials}, we expand the $\theta$-function expression at several (computer generated) random points for the gauge and flavor fugacities.}.

\subsection{Theories without charged matter}
\label{sub:matter-less}
Theories without matter necessarily exhibit trivial flavor symmetry $F$: the sum over $\lambda$ is absent for these theories, such that 
\begin{equation}
    \ER=\xi^{n,G} \,.
\end{equation}
The absence of the left-moving fermionic bundle in the non-linear sigma model description of the worldsheet theory significantly simplifies the structure of the elliptic genus.  

Indeed, for the matterless models $(D_4)_4,\,(F_4)_5,\,(E_6)_6,\,(E_7)_8$, we find 
\begin{equation}
    \xi^{n,G}= \frac{1}{\eta(q)}\sum_n\hat \chi_{n\theta}\sum_{m=0}^2a_m\left(v^{-b_{n,m}+\kappa/2}q^{-\frac{(b_{n,m}-\kappa/2)^2}{4\kappa}}-(-1)^s v^{b_{n,m}+\kappa/2}q^{-\frac{(b_{n,m}+\kappa/2)^2}{4\kappa}}
    \right) \,,
    \label{eq:massless_result}
\end{equation}
where 
\begin{equation*}
    b_{n,m}=2n+\kappa m-(\kappa-2),\quad \kappa=h^\vee_G-n\,,
\end{equation*}
and $a,s$ are given in table \ref{tab:tab_as}.
The $(D_4)_4$ result was already pointed out in \cite{DelZotto:2018tcj}.
\begin{table}[]
    \centering
    \begin{tabular}{c|cccc}
         & $a_0$&$a_1$&$a_2$&$s$ \\\hline
        $(D_4)_4$ & 1&2&1&0\\
        $(F_4)_5$ & -1&0&1&1\\
        $(E_6)_6 $& 1&2&1&0\\
        $(E_7)_8$&1&2&1&0
    \end{tabular}
    \caption{Coefficients and relative sign in expansion of the elliptic genus of non-Higgsable models in affine characters}
    \label{tab:tab_as}
\end{table}
We note that the operator $\mathcal F_\kappa$ simply permutes the two terms in parentheses and the sign $(-1)^s$ determines whether $\ER$ is periodic or anti-periodic under this transformation. 

Note that for $(D_4)_4$, the UV theory was derived in \cite{Haghighat:2014vxa} and $\ER$ as a function of the gauge and flavor fugacities determined in terms of $\theta$-functions. Obtaining the result \eqref{eq:massless_result} for this theory hence does not require the algorithm presented in section \ref{sec:algorithm}. 

\subsection{$(E_7)_7$ and $(B_4)_4$}
\label{subsec:E7_7} 
$\mathbf{(E_7)_7}$

$(E_7)_7$ is the only theory with matter and without flavor group. We find that the elliptic genus can be written as 

\begin{align}
\footnotesize
\begin{split}
        \xi^{7,E_7}=\frac{1}{\eta(q)}\Bigg(\sum_n \hat \chi_{n\theta}\sum_{m=0}^2a_m v^{-b^+_{n-1,m}+\kappa/2}q^{-\frac{(b^+_{n-1,m}-\kappa/2)^2}{4\kappa}}+\sum_{n}\hat \chi_{n\theta+(0000001)}\sum_{m=0}^2a_m v^{ b^+_{n,m}+\kappa/2}q^{-\frac{(b^+_{n,m}+\kappa/2)^2}{4\kappa}}\\
        -\sum_n \hat \chi_{n\theta+(0000010)}\sum_{m=0}^2a_m v^{-b^-_{n,m}+\kappa/2}q^{-\frac{(b^-_{n,m}-\kappa/2)^2}{4\kappa}}-\sum_{n}\hat \chi_{n\theta+(0100000)}\sum_{m=0}^2a_m v^{ b^-_{n+1,m}+\kappa/2}q^{-\frac{(b^-_{n+1,m}+\kappa/2)^2}{4\kappa}}\\
        -\hat \chi_{(0000100)}\sum_{m=0}^2a_m v^{b^-_{0,m}+\kappa/2}q^{-\frac{(b^-_{0,m}+\kappa/2)^2}{4\kappa}}+\hat \chi_{(0001000)}\sum_{m=0}^2a_m v^{b^+_{-1,m}+\kappa/2}q^{-\frac{(b^+_{-1,m}+\kappa/2)^2}{4\kappa}}\Bigg)\,,
\end{split}
\end{align}

where 
\begin{align}
    b^\pm_{m,n}=(-1)^m\left(2n+\kappa m-(\kappa-2)\pm \frac{1}{2}\right),\, a_0=1,\, a_1=-2,\, a_2=1\,.
\end{align}

Due to the presence of matter, we expect $\mathcal{F}_\kappa$ to not act trivially on $\xi$; however, as there is no flavor group, $\mathcal F_\kappa$ cannot act as a permutation of the flavor weights. We observe that $\mathcal F_\kappa$ permutes the contributions of the different characters. In terms of the polynomials in equation \eqref{eq:poly_multiplying_omega},
\begin{equation}
    \mathcal{F}_\kappa (p_\omega)(y,v)=p_{\sigma(\omega)}(y,v)\,,
\end{equation}
where $\sigma^2=1$ and 
\begin{align*}
    \sigma(n\theta)&=(n+1)\theta+(0000001),\quad n\geq 0\,,\\
    \sigma(0)&=(0001000)\,,\\
    \sigma(n\theta+(0100000))&=(n+1)\theta+(0000010)\,, \quad n\geq 0\\
    \sigma(0000010)&=(0000100) \,.
\end{align*}

$\mathbf{(B_4)_4}$

For $(B_4)_4$, the flavor group is $A_1$ at level $1$. We observe that the $\xi$ functions can be written as

\begin{align}
\begin{split}
        \xi^{4,B_4}_{(0)}&=\frac{1}{\eta(q)}\Bigg(\sum_{n>0} \hat \chi_{n\theta}\sum_{m=0}^3a_m v^{b^-_{n,m}+\kappa/2}q^{-\frac{(b^-_{n,m}+\kappa/2)^2}{4\kappa}}+\sum_{n}\hat \chi_{n\theta+(1000)}\sum_{m=0}^3a_m v^{ -b^+_{n,m}+\kappa/2}q^{-\frac{(b^+_{n,m}-\kappa/2)^2}{4\kappa}}\\&
        +\hat \chi_{(0000)}\sum_{m=2,3}a_m v^{b^-_{0,m}+\kappa/2}q^{-\frac{(b^-_{0,m}+\kappa/2)^2}{4\kappa}}+\hat \chi_{(2000)}\sum_{m=0,1}a_m v^{b^-_{0,m}+\kappa/2}q^{-\frac{(b^-_{0,m}+\kappa/2)^2}{4\kappa}}\Bigg)\,,
\end{split}\\
\xi^{4,B_4}_{(1)}&=-\mathcal{F}_\kappa(\xi^{4,B_4}_{0})
\end{align}

with

\begin{equation*}
    b_{n,m}^\pm=(-1)^m\left(2n+\kappa m+(\kappa-2) -6 \pm \frac{1}{2}\right), \, a_0=1,\,a_1=-1,\,a_2=-1,\,a_3=1\,.
\end{equation*}

\subsection{The $C_r$ branch of the E-string Higgsing tree}
\label{sub:E-string results}
The elliptic genera of the theories of the $E$-string Higgsing tree ($n=1$) with gauge group $C_r$ where given in \cite{DelZotto:2018tcj} for $r=1,2$ . We conjecture a general result for arbitrary $r>1$:
\begin{align} \label{eq:Cr}
    \eta\, \xi_0^{C_r}&=\hat\chi_0+\hat\chi_{\Lambda_2}+\sum_{k\geq 1}  \hat\chi_{2k\Lambda_1} q^{-\frac{(2k)^2}{4\kappa}}(v^{2k}+v^{-2k})\,,\\
    \eta \,\xi_v^{C_r}&= - \sum_{k\geq 1}  \hat\chi_{(2k-1)\Lambda_1} q^{-\frac{(2k-1)^2}{4\kappa}} (v^{2k-1}+v^{-(2k-1)})\,,\nonumber\\
    \xi_s^{C_r}&=\mathcal{F}_\kappa\xi_1^{C_r}\,, \nonumber\\
    \xi_c^{C_r}&=\mathcal{F}_\kappa\xi_v^{C_r}\,. \nonumber
\end{align}
Recall that the operators $\mathcal{F}_\kappa$ were introduced in equation \eqref{eq:def_F}. Also, the flavor symmetry of the $(C_r)_1$ theory is $D_{8+2r}$ at level 1. Only the representations with fundamental weight associated to the two extremities of the affine Dynkin diagram contribute; we retain the same name for these as in the case of $D_4$. The result at $r=1$ takes a slightly different form:
\begin{align} \label{eq:C1}
    \eta\, \xi_0^{C_1}&=\sum_{k\geq 1}  \hat\chi_{2k\Lambda_1} \sum_{m=-k}^k q^{-m^2}v^{2m}\,,\\
    \eta \,\xi_v^{C_1}&= - \sum_{k\geq 1}  \hat\chi_{(2k-1)\Lambda_1} \sum_{m=-k}^k q^{-\frac{(2m+1)^2}{4}}v^{-1-2m}, \nonumber\\
    \xi_s^{C_1}&=\mathcal{F}_1\xi_1^{C_1}\,,\nonumber\\
    \xi_c^{C_1}&=\mathcal{F}_1\xi_v^{C_1}\,, \nonumber
\end{align}
where we have explicitly substituted $\kappa=1$.

Note that just as for the $(D_4)_4$ theory, a UV description of the $(C_r)_1$ theories is known; the elliptic genera can hence be computed exactly in terms of $\theta$-functions \cite{Kim:2014dza}. The results \eqref{eq:Cr} and \eqref{eq:C1} were consequently obtained without recourse to the algorithm presented in section \ref{sec:algorithm}. The fact that the coefficients $c^\lambda_{\omega,k}$ all equal $\pm 1$ thus provides additional evidence for the conjectured form \eqref{eq:affineAnsatzIntro} of the elliptic genus which lies at the heart of this work. 

We remark that for these theories, we can combine the knowledge of $\ER$ in terms of $\theta$-functions with the affine ansatz \eqref{eq:affine_G_expansion} to derive explicit formulae for (some) level $-1$ characters of $C_r$.

Let us consider $r>1$, as only a single affine character contributes at each power of $v$ in this case (except for $v^0$ in $\xi_0$). The functions $\xi_0,\,\xi_v$ in terms of $\theta$-functions are given by \cite{Kim:2014dza,DelZotto:2018tcj}
\begin{equation}
     \xi_{0/v}^{C_r}=\frac{1}{2}\left(\prod_{\substack{1\leq i\leq r\\s=\pm}}^r\frac{\eta}{\theta_3(v (X_i^{C_1})^s)}\pm\prod_{\substack{1\leq i\leq r\\s=\pm}}^r\frac{\eta}{\theta_4(v (X_i^{C_1})^s)}\right)\,,
     \label{eq:xi_n=1_theta}
\end{equation}
where the $+$ sign corresponds to $0$ and the $-$ to $v$. We then have that
\begin{align}
\hat\chi_{\ell\Lambda_1}&=q^\frac{\ell^2}{4r}\left[\frac{1}{2}\left((-1)^\ell\prod_{\substack{1\leq i\leq r\\s=\pm}}^r\frac{\eta}{\theta_3(v (X_i^{C_1})^s)}+\prod_{\substack{1\leq i\leq r\\s=\pm}}^r\frac{\eta}{\theta_4(v (X_i^{C_1})^s)}\right)\right]_{v^{\ell}}\,,
\end{align}
where $[\cdot]_{v^\ell}$ signifies the order $v^\ell$ term of $[\cdot]$.\footnote{We could of course equally well choose the term $v^{-\ell}$.} This expression can be further simplified using the explicit forms or the modular transformation properties of the $\theta$- functions to give
\begin{equation}
    \hat\chi_{\ell\Lambda_1}=q^\frac{\ell^2}{4r}\left[\prod_{\substack{1\leq i\leq r\\s=\pm}}^r\frac{\eta}{\theta_4(v (X_i^{C_1})^s)}\right]_{v^{\ell}}\,.
\end{equation}
To our knowledge, this expression for level $-1$ characters of $C_r$ has not appeared previously in the literature. It would be interesting to compare it to a recent result by Kac and Wakimoto \cite{Kac2018} on these characters.

\section{Relations along the Higgsing tree} \label{s:Higgs_tree_relations}
When explicit results for the elliptic genera of a pair of theories related by Higgsing are known in terms of $\theta$-functions, it has been found in the literature \cite{Okuda_2012,DelZotto:2018tcj,Gu:2020fem} that the elliptic genus of the Higgsed theory can be obtained by a specialization of the parameters of the parent theory. Notably, some flavor fugacities of the parent theory must be replaced by the fugacity $v$ of the $SU(2)_v$ symmetry. In section \ref{ss:specializing}, we determine this specialization map for a number of theories, notably for multiple theories for which no expressions for the elliptic genus in terms of $\theta$-functions is known. The fact that also in these cases, the specialization map is of the expected form provides additional evidence for the various assumptions that enter in our derivation of the affine presentation of $\ER$.

In section \ref{sub:constrainguptree}, we revisit an idea of \cite{Duan:2020imo}, which demonstrated that the computation of elliptic genera can sometimes be simplified by imposing a larger Weyl symmetry, one which occurs ``further up the tree'' from the theory $G_n$ in question, in choosing the ring of Jacobi forms than the one suggested by the group $G$. We will consequentially consider the problem of expanding the elliptic genera with characters of a Lie algebra that correspond to a possible unHiggsing up the tree. Note that such an expansion is necessarily possible as long as additional $q$ dependence outside that occurring in the group and flavor characters is permitted: it imposes a larger (finite) Weyl symmetry on the $m_G$ dependence of $\ER$, which is permissible following the results of \cite{Duan:2020imo}. However, we will see that the constrained form of the ansatz \eqref{eq:affineAnsatzIntro} no longer holds for this larger Lie algebra. The possibility of unHiggsing can however be used to constrain the representations $\omega$ occurring in the ansatz \eqref{eq:affineAnsatzIntro} further. We will illustrate these points using the example of the Higgsing $(F_4)_4\to (D_4)_4$.

\subsection{Higgsing via specialization of fugacities} \label{ss:specializing}

Given a theory with gauge and flavor groups $G,\,F$ and a Higgsing down the Higgsing tree to a theory with gauge and flavor groups $G',\,F'$, we find that there are two maps \cite{Okuda_2012,Gu:2020fem,DelZotto:2018tcj}
\begin{equation} \label{eq:specialization_maps}
    \iota_G:\mathfrak h'\to\mathfrak h,\quad \text{and }\quad \iota_F:\mathfrak f'\oplus \mathfrak h_v\to\mathfrak{f}\,,
\end{equation}
with $\mathfrak h',\,\mathfrak h,\,\mathfrak f',\,\mathfrak f,$ and $\mathfrak h_v$ denoting the Cartans of the Lie algebras associated to $G',\,G,\,F',\,F, $ and $SU(2)_v$ respectively, such that
\begin{equation}
   \E^{(G',F')}(m_G',m_F',v,q)=\iota^*\left(\E^{(G,F)}\right)(m_G',m_F',v,q):=\E^{(G,F)}(\iota_G (m_G'),\iota_F(m_F'),v,q)\,.
\end{equation}
The map $\iota_G$ was described for most gauge groups in \cite{Duan:2020imo}, where the same question was studied from the point of view of the modular ansatz for $\ER$. The map $\iota_F$ is given by the same transformation for the flavor part $\mathfrak f$, and an inclusion of $\mathfrak{h}_v$ in an orthogonal direction. 

We will describe these transformations explicitly in several examples in the following. We give the functions $\iota^*$ acting in either the exponentiated fugacities of the Euclidean lattice $X_i=e^{2\pi i (e_i,m_G)}$ or in terms of the exponentiated fugacities corresponding to the Lie algebra roots $Q_i=e^{2\pi i(\alpha_i,m_G)}$. We follow the same normalization conventions outlined in appendix C of \cite{Duan:2020imo}.


{\bf $\mathbf C_r$ tower}

The only theories with $C_r$ gauge symmetry appear on a branch of the E-string Higgsing tree. The specialization maps \eqref{eq:specialization_maps} can be read off \cite{DelZotto:2018tcj} from the explicit expression of the elliptic genera in terms of $\theta$-functions \cite{Kim:2014dza}: the Cartan algebra of $C_r$ is mapped to the hyperplane $x_{r+1}=0$ of the Euclidean space in which the Cartan algebra of $C_{r+1}$ is embedded. For the flavor group, one identifies the Cartan algebra of $D_{8+2r}$ with the co-dimension 2 space $x_{8+2r+1}=x_{8+2r+2}=0$ of the corresponding higher-dimensional Euclidean space. The fugacities corresponding to these two directions are then replaced by $v$. Explicitly,
\begin{equation}
   \iota_G^*: X_i^{C_{r+1}}\mapsto\left\{\begin{array}{cc}
        X_i^{C_r} & i=1,\dots, r  \\
        1 & i=r+1
    \end{array}\right.\!,\, \iota_F^*: X_i^{D_{8+2(r+1)}}\mapsto \left\{\begin{array}{cc}
        X_i^{D_{8+2r}} & i=1,\dots, 8+2r  \\
        v & i=8+2r+1, 8+2r+2
    \end{array}\right..
\end{equation}

{\bf $\mathbf{B/D\sim SO(N)}$ towers}

The Higgsing trees at $n=1,2,3,4$ exhibit a (finite or infinite) branch of alternating $B$ and $D$ theories.

We study the $n=4$ case for which the exact elliptic genus is known in terms of $\theta$-functions \cite{Kim:2014dza,DelZotto:2018tcj}. From this expressions one can get the fugacity transformation for the different Higgsings. For the $B_r\to D_r$ Higgsing we identify the two Cartan sub-algebras through their embedding in Euclidean space and the $C_{r-3}\to C_{r-4}$ flavor fugacities as in the previous example, except that we set the last fugacity to $v$. Explicitly,
\begin{equation}
    \iota_G^*:X_i^{B_{r}}\mapsto X_i^{D_r},\quad \iota_F^*:X_i^{C_{r-3}}\mapsto \left\{\begin{array}{cc}
        X_i^{C_{r-4}} & i=1,\dots, r-4  \\
        v & i=r-3
    \end{array}\right.\,.
\end{equation}

For the Higgsing $D_{r+1}\to B_{r}$, the flavor fugacities transform in the same fashion, while the $(r+1)^{st}$ $D_{r+1}$ fugacity must be set to 1:
\begin{equation}
    \iota_G^*: X_i^{D_{r+1}}\mapsto\left\{\begin{array}{cc}
        X_i^{B_r} & i=1,\dots, r  \\
        1 & i=r+1,
        \end{array}\right.
        \quad 
        \iota_F^*:X_i^{C_{r-2}}\mapsto \left\{\begin{array}{cc}
        X_i^{C_{r-3}} & i=1,\dots, r-3  \\
        v & i=r-2
    \end{array}\right.\,.
\end{equation}

{$\mathbf{F_4\to D_4}$}

\label{sub:F4toD4Qtransformation}
The Higgsing trees with $n=1,2,3,4$ each exhibit a Higgsing from a theory with gauge group $F_4$ to one with gauge group $D_4$. Only for the $(D_4)_4$ theory is the elliptic genus known exactly in terms of $\theta$-functions.

We study the $n=4$ case. The transformation of the gauge fugacities is the same as in \cite{Duan:2020imo}:
\begin{equation}
        Q_1^{F_4} \mapsto Q_2^{D_4} \,,\,\, Q_2^{F_4} \mapsto Q_1^{D_4} \,,\,\, Q_3^{F_4} \mapsto \sqrt{\frac{Q_3^{D_4}}{Q_1^{D_4}}}  \,,\,\, Q_4^{F_4} \mapsto \sqrt{\frac{Q_4^{D_4}}{Q_3^{D_4}}}  \,, 
        \label{eq:F4toD4Qtransformation}
    \end{equation}
while for the flavor fugacity we have
\begin{equation*}
    X^{C_1}\mapsto v.
\end{equation*}
We will revisit this example in section \ref{sub:constrainguptree}.

{$\mathbf{E_6 \to F_4}$}

For the $n=5$ tree, we consider the elliptic genera of $(F_4)_5$ and $(E_6)_5$. To find the appropriate replacement of the gauge parameters, we use the transformation \eqref{eq:F4toD4Qtransformation} to identify the $D_4$ and $F_4$ root lattices. Then, the $D_4$ root lattice can be embedded in the root lattice of $E_6$ simply by identifying the $D_4$ Euclidean space with the space $x_5=0,x_6=0$ in the Euclidean space of $E_6$. This gives the $\iota_G^*$ transformation 
\begin{equation}
    \begin{array}{ccc}
 Q_1^{E_6}\to \frac{1}{Q_3^{F_4} Q_4^{F_4}}\,,&Q_2^{E_6}\to \frac{1}{Q_2^{F_4}}\,,&Q_3^{E_6}\to \frac{1}{Q_1^{F_4}},
     \\Q_4^{E_6}\to
   \frac{1}{Q_2^{F_4} \left(Q_3^{F_4}\right)^2}\,,&Q_5^{E_6}\to \frac{1}{Q_4^{F_4}}\,,&Q_6^{E_6}\to \left(Q_1^{F_4}\right)^2
   \left(Q_2^{F_4}\right)^3 \left(Q_3^{F_4}\right)^4 \left(Q_4^{F_4}\right)^2\,.
    \end{array}
    \label{eq:E6toF5}
\end{equation}
The flavor fugacity transformation is simply $\iota_F^*:Q^{U_1}\mapsto v$. 

Recall that we can fix the functions $\xi$ for $(E_6)_5$ only up to a Dynkin symmetry, as explained at the end of section \ref{ss:strategy}. We unfortunately cannot resolve this ambiguity using the Higgsing, as the finite representations of $E_6$ related by the Dynkin symmetry have the same image under the transformation \eqref{eq:E6toF5},
\begin{equation}
   \iota_G^* \chi^{E6}_{(a_1,a_2,a_3,a_4,a_5,a_6)}=\iota_G^* \chi^{E6}_{(a_5,a_4,a_3,a_2,a_1,a_6)}
\end{equation}

\subsection{Enhanced Weyl symmetry $\nRightarrow$ enhanced affine symmetry}
\label{sub:constrainguptree}
In \cite{Duan:2020imo}, a variety of circumstances were found under which the Weyl symmetry $\cW_G$ of a $G_n$ theory is enhanced.\footnote{In many cases, we also observe an enhancement of the coroot translational invariance of the elliptic genus to finer lattices. This is explained by the absence of certain representations in the charged spectrum or delicate cancellation between matter and gauge contributions \cite{Kashani-Poor:2019jyo}.} The possible enhancements include $\cW_{D_4}\to \cW_{F_4}$, $\cW_{B_4}\to \cW_{F_4} $, $\cW_{A_2}\to \cW_{G_2}$, and $\cW_{D_n}\to \cW_{B_n}$. This observation was put to good use in choosing to expand the elliptic genus in terms of Jacobi forms invariant under the enhanced Weyl symmetry. It is thus natural to ask whether in the context of our affine ansatz \eqref{eq:affineAnsatzIntro}, enhanced Weyl symmetry implies a simplified ansatz in terms of affine characters of the Lie algebra associated to it. This sadly does not appear to be the case. We will use the example $(F_4)_4 \rightarrow (D_4)_4$ which already appeared above to illustrate this point.

Let us thus look at this example in somewhat more detail. The  transition $\iota$ in this case is implemented by the identification of the $D_4$ and $F_4$ coweight lattices and the transformation of the unique flavor fugacity to $v$ as explained in subsection \ref{sub:F4toD4Qtransformation}. At leading order in $q$, the functions $\xi$ and the corresponding flavor characters (specialized to the value $v$) are given by
\begin{equation}
    \begin{array}{ccc}
     \lambda& \hat\chi_\lambda(v) &\eta\xi  \\\hline
    (0) &1&-v^7 \chi^{F_4}_{(0000)}+v^{11}\sum v^{2j}\chi^{F_4}_{(j100)}\\
     (1) &v^{-1}+v& -v^{10}\sum_j v^{2j}\chi^{F_4}_{(j010)}\\
     (2) &v^{-2}+1+v^2& v^{9}\sum_j v^{2j}\chi^{F_4}_{(j001)}\\
     (3) &v^{-3}+v^{-1}+v+v^3&-v^{8}\sum_j v^{2j}\chi^{F_4}_{(j000)}
\end{array}
\end{equation}
To recover the elliptic genus of $(D_4)_4$, we must multiply the two columns, add the resulting rows, and express the $F_4$ characters in terms of $D_4$ characters. This yields
\begin{equation} \label{eq:leadingD4inF4}
\begin{aligned}
        &-v^5\chi^{F_4}_{(0000)}-v^7(\chi^{F_4}_{(1000)}-\chi^{F_4}_{(0001)}+2\chi^{F_4}_{(0000)})\\
        &-v^9(\chi^{F_4}_{(2000)}-\chi^{F_4}_{(1001)}+\chi^{F_4}_{(1000)}+\chi^{F_4}_{(0010)}-\chi^{F_4}_{(0001)}+\chi^{F_4}_{(0000)})\\
        &-\sum_{j\geq 3} v^{5+2j}(\chi^{F_4}_{(j000)}-\chi^{F_4}_{(j-1,001)}+\chi^{F_4}_{(j-1,000)}+\chi^{F_4}_{(j-2,010)}-\chi^{F_4}_{(j-2,001)}\\&+\chi^{F_4}_{(j-2000)}+\chi^{F_4}_{(j
        -3,010)}-\chi^{F_4}_{(j-3,001)}+\chi^{F_4}_{(j-3,000)})\\
        &=\sum_jv^{5+2j}\chi^{D_4}_{0j00}\,,
\end{aligned}\end{equation}
where in the last line we have implemented \eqref{eq:F4toD4Qtransformation}. We draw two conclusions from this calculation: firstly, that it is possible to write the $(D_4)_4$ elliptic genus in terms of affine characters of $F_4$: \eqref{eq:leadingD4inF4} demonstrates this for the finite characters at leading order in $q$. Replacing all characters by their affine counterparts preserves the equality at leading order in $q$. We can correct the expression at the next order in $q$ by subtracting affine characters of $F_4$ that are induced but do not occur in the $(D_4)_4$ elliptic genus at this level, and by adding in affine characters of $D_4$ that do not arise by affinizing the $F_4$ characters at lower level. For this procedure to work, it is of course crucial that the gauge fugacities of the $(D_4)_4$ elliptic genus exhibit $\cW_{F_4}$ symmetry. This was observed in \cite{Duan:2020imo}, and is manifest in the affine expansion, as the sum over $\omega$ is over Dynkin symmetric representations $(0j00)$ only (at each order in $q$, the finite representations contributing to $\hat \chi^{D_4}_{(0j00)}$ occur in Dynkin symmetric combinations). Note that we have permitted ourselves to introduce $q$ dependence beyond that carried by the characters. The second conclusion we draw from the result \eqref{eq:leadingD4inF4} is that expressing the leading contribution to the $(D_4)_4$ elliptic genus in terms of $F_4$ characters has rendered the result more cumbersome. In particular, the structure \eqref{eq:affineAnsatzIntro} is not preserved.

The source of this complication is of course the additional $v$ dependence contributed by $\hat \chi_\lambda(v)$. Consider e.g. the $F_4$ character $\chi^{F_4}_{(j000)}$. It appears at leading order in $q$ in the $(F_4)_4$ elliptic genus because 
\begin{equation*}
    x_0((3),(j000),8+2j)=\casimirR\,.
\end{equation*}
However, in the $(D_4)_4$ elliptic genus, there is a contribution from the same character at orders $8+2j+\delta$ where $\delta={-3,-1,1,3}$ (the powers of $v$ appearing in $\chi^{C_1}_{(3)}$). Therefore, any prescription for expressing $\E^{D_4}$ in terms of $F_4$ characters would need to predict a term at order $k$ in $v$ for every $F_4$ dominant weight $\omega $ that satisfies 
\begin{equation*}
    x_0((3),\omega,k-\delta)=\casimirR \quad \text{ for any choice of } \delta={-3,-1,1,3}\,,
\end{equation*}
or that satisfies the analogous equation for any of the other $\lambda$'s. At higher order in $q$ the problem gets amplified because there are $q$-dependent contributions coming from $\hat \chi_\lambda(v)$. 

    To end on a positive note, we point out that expressing the characters of the Higgsed gauge group in terms of the characters of the parent gauge group does teach us something about the parent theory: which characters must occur. Thus, the relation \eqref{eq:leadingD4inF4} tells us that characters of the form $(j000),\,(j100),\,(j010),\,(j001)$ must appear in the $(F_4)_4$ elliptic genus, to make the specialization to the $(D_4)_4$ elliptic genus possible.

    \section{Conclusions} \label{s:conclusions}
    We have put forth in this paper a conjecture on the form of the elliptic genus of the non-critical strings of 6d field theories obtained from F-theory compactifications on elliptically fibered Calabi-Yau manifolds over Hirzebruch bases, given in equations \eqref{eq:affineERintro}, \eqref{eq:affineAnsatzIntro} and \eqref{eq:constraintIntro}, and provided ample evidence for its validity. Our results match expressions for the elliptic genus when these are known, and fit into specialization sequences in Higgsing trees containing known results. 

    It would be interesting to extend our ansatz beyond the class of theories discussed here, e.g. to the class F-theory compactifications on elliptic fibrations without section \cite{Cota:2019cjx} or theories obtained via twisted compactification \cite{Duan:2021ges}. Both constructions lead to elliptic genera involving Jacobi forms of congruence subgroups of $SL(2,\IZ)$. Also, studying the elliptic genera of multiple strings is a natural next step.

    Already for the non-critical strings whose elliptic genera we explored in this paper, an important challenge remains to be met: to explain our ansatz and the results of our computation from the worldsheet theory of the non-critical string. What is the origin of the non-integrable affine symmetry of $G$? More ambitiously still, can the closed form results presented in section \ref{s:closed_form_results} and the polynomials $p_\omega^\lambda$ introduced in equation \eqref{eq:poly_multiplying_omega} and computed for many examples in appendix \ref{a:polynomials} be computed from the vantage point of the worldsheet theory? We hope to return to these question elsewhere.

    \section*{Acknowledgements}

    We would like to thank Michele Del Zotto, Gugliemo Lockhart, Lionel Mason, Noppadol Mekareeya, Thorsten Schimannek and Timo Weigand for useful conversations. Special thanks to Balt van Rees for sharing the Mathematica code from the project \cite{Beem:2013sza} for computing negative level characters of $D_4$ with us. This helped speed up the development of our own SageMath code for arbitrary simple Lie algebras substantially.

    A.K.\ acknowledges support under ANR grant ANR-21-CE31-0021.

\appendix

    \section{Elliptic genera} \label{a:elliptic_genera}
    
    An $N=1$ right-moving supersymmetry $Q_R$ in two dimensions together with a right-moving fermion charge operator $F_R$ permits defining a trace \cite{Witten:1982df}
    \begin{equation}
        \tr_{\cH} (-1)^{F_R} \bar{q}^{H_R}
    \end{equation}
    which receives contributions only from states at $H_R=0$. When the kernel of $H_R$ is infinite dimensional, additional operators must be included in the trace to render it well-defined, 
    \begin{equation} \label{eq:flavored_trace}
        \tr_\cH (-1)^{F_R} X q^{H_L} \bar{q}^{H_R} \,.
    \end{equation}
    To maintain the property that the trace receives contributions only at $H_R =0$, two states $\psi$ and $Q_R \psi $ must have the same $X$ eigenvalue, i.e. only operators $X$ that commute with $Q_R$ are permissible insertions.
    
    An additional right-moving supersymmetry $\bar{Q}_R$ yielding an $\cN=2$ algebra
    \begin{equation} \label{eq:N2}
        Q_R^2 = \bar{Q}_R^2 = 0 \,, \quad \{Q_R , \bar{Q}_R\} = 2H_R
    \end{equation}
    permits interpreting the elements of the kernel of $H_R$ as representatives of the cohomology of $\bar{Q}_R$. Central to this argument is that $\bar{Q}_R$ be the adjoint operator to $Q_R$, such that\footnote{In two dimensions, the Weyl and Majorana conditions can simultaneously be imposed on spinors. Consequently, supercharges can be chosen to be chiral and hermitian \cite{Polchinski:1998rq}. Multiple such supercharges satisfy the algebra
    \begin{eqnarray} \label{eq:Majorana_Weyl_supercharges}
        \{Q_L^A , Q_L^B \} &=& \delta^{AB} (P^0 - P^1) = \delta^{AB} H_L \,,\\
        \{Q_R^A , Q_R^B \} &=& \delta^{AB} (P^0 + P^1) = \delta^{AB} H_R \,, \nn \\
        \{Q_L^A , Q_R^B \} &=& Z^{AB}\,. \nn
    \end{eqnarray}
    For the purposes of defining the elliptic genus, it is more convenient to work with a pair of charges that are adjoint to each other, to mimic the behavior of the $\bar{\partial}$ and $\bar{\partial}^\dagger$ operators in Hodge theory. These can be defined as
    \begin{equation}
        Q_R = Q_R^1 + i Q_R^2 \,, \quad Q_R^\dagger = Q_R^1 - i Q_R^2
    \end{equation}
    with anti-commutation relations
    \begin{equation} \label{eq:adjoint_supercharges}
        Q_R^2 = (Q_R^\dagger)^2 = 0 \,, \quad \{Q_R, Q_R^\dagger\} = 2 H_R \,.
    \end{equation}
    Note that the $SO(2)$ symmetry of the right moving sector in the presentation \eqref{eq:Majorana_Weyl_supercharges} (restricted to a pair of charges) is realized as a $U(1)$ symmetry of the algebra \eqref{eq:adjoint_supercharges}.}
    \begin{equation}
        \langle \{Q_R , \bar{Q}_R \} \psi , \psi \rangle = || Q_R \psi ||^2 + || \bar{Q}_R \psi ||^2 \,.
    \end{equation}
    If $\psi \in \ker H_R$, it is therefore both $Q_R$ and $\bar{Q}_R$ closed. On the other hand, $\bar{Q}_R \psi \in \ker H_R \Rightarrow \bar{Q}_R \psi = 0$ by a similar argument. Acting by $\bar{Q}_R$ hence either annihilates the state or maps out of the kernel of $H_R$. In both cases, the image does not contribute to the trace \eqref{eq:flavored_trace}; it is therefore only sensitive to the $\bar{Q}_R$ cohomology. In the context of $\cN=2$ theories, the trace \eqref{eq:flavored_trace} is referred to as the elliptic genus \cite{Witten:1986bf, Witten:1993jg}.
    
    This cohomological underpinning gives the elliptic genus its stability under sufficiently mild variations of the Lagrangian.

    In the case of $\cN=2$ superconformal symmetry, the anti-commutation \eqref{eq:N2} holds in the Ramond sector with $H_R = \bar L_0 - \frac{c_R}{24}$, while in the NS sector, it becomes
    \begin{equation}
        \{Q_R, \bar Q_R \} = \bar L_0 + \frac{1}{2} \bar J_0 \,, \label{eq:anticommutation_NS}
    \end{equation}
    with $\bar J_0$ the $U(1)$ charge generator of the $\cN=2$ superconformal algebra. All the consideration which were formulated above thus go through for $H_R$ replaced by the RHS of the relation \eqref{eq:anticommutation_NS}.

    Evaluation of the elliptic genus in Landau-Ginzburg models, by deforming away the potential, boils down to multiplying the universal contributions from each multiplet, followed, in the case of gauge theories, by an integration over holonomies \cite{Benini:2013a, Benini:2013xpa}.

\section{Computing affine characters at negative level via Kazhdan-Lusztig polynomials} \label{a:Kazhdan-Lusztig}

The character of a representation $R$ of an affine Lie algebra $\affineg$ is given by 
\begin{equation*}
    \cha R=\sum_{\mu\in \LambdaW} m_R(\mu) e^{2\pi i\mu}
\end{equation*}
where the sum is over the weight lattice $\LambdaW$ and $m_R(\mu)$ denotes the multiplicity of the weight $\mu$ in the representation.

The characters $\hat \chi_\lambda$ for the highest weight representation $L_\lambda$ that appear in the main text differ from $\cha L_\lambda$ by a shift in the powers of $q$ given by the ground state energy level of the Wess-Zumino-Witten model associated to $G$:
\begin{equation}
    \hat \chi_\lambda =q^{-\frac{c}{24}+h_\lambda}\cha L_\lambda\,,
\end{equation}
with 
\begin{equation*}
    c=\frac{\dim(G)k}{h_G^\vee+k}\,, \quad h_\omega^G=\frac{\left<\lambda,\lambda+2\rho\right>}{h_G^\vee+k}\,,
\end{equation*}
where $k$ is the level of $\lambda$, $h^\vee_G$ the dual Coxeter number of $G$, $\rho$ the Weyl vector, and the inner product $\langle \cdot , \cdot \rangle$ is normalized to $2$ for long roots.

For positive level representations of affine Lie algebras, the Weyl-Kac formula
\begin{equation}
    \cha L_\lambda = \sum_{\omega\in W}\text{sign}(w)\cha M_{\omega \cdot \lambda}
    \label{eq:weyl-kac}
\end{equation}
gives the character of an irreducible module $L_\lambda$ of highest weight $\lambda$ in terms of Verma module characters $M_\mu$, whose characters can easily be computed as
\begin{equation}
    \cha M_\mu=\frac{e^\mu}{\prod_{\alpha\in \Delta_+}(1-e^\alpha)^{r_\alpha}}, \quad r_\alpha =\left\{\begin{array}{cc}
         \text{Rank}(\mathfrak{g})& \text{ if } \alpha\in \mathbb N\delta  \\
         1& \text{otherwise}
    \end{array}\right.\,.
\end{equation}
The representations of the flavor group which appear in the ansatz \eqref{eq:affineERintro} are at positive level. There exist efficient algorithms to compute such characters: Freudenthal's formula e.g. computes $m_\lambda(\mu)$, the multiplicity $m_\lambda(\mu)$ of the weight $\mu$ in the highest weight representation $L(\lambda)$ as
    \begin{equation} \label{eq:Freudenthal}
        (||\lambda+\rho||^2-||\mu+\rho||^2)m_\lambda(\mu)=2\sum_{\alpha \in \Delta^+}\sum_{j\geq 1}(\lambda+j\alpha,\alpha)m_\lambda(\lambda+j\alpha)\,,
    \end{equation}
    with $\Delta^+$ the set of positive roots.

The theory underlying characters of negative level representations is more intricate. In particular, the representation theory is no longer invariant under the action of the affine Weyl group $\affineWeyl$ (though the symmetry under the action of the finite Weyl group $\finiteWeyl$ is preserved if all finite Dynkin labels of the highest weight are non-negative). The Kazhdan-Lusztig conjecture states that for such representations, the Weyl-Kac like relation between the sought after character and Verma module characters is 
\begin{equation}
    \cha L_{\lambda}=\sum_{w\leq w'}m_{w,w'}\,\cha M_{w'\cdot \Lambda}\,.
    \label{eq:klchar}
\end{equation}
The ingredients in this formula are the following: $\Lambda$ is the unique weight in the Weyl orbit $\affineWeyl \cdot \lambda$ under the dotted action
\begin{equation}
    \affineWeyl \cdot \lambda= \{ w(\lambda+\rho)-\rho \, | \, w \in \affineWeyl \} 
\end{equation}
such that $\Lambda+\rho$ is a dominant weight. The affine Weyl group element $w \in \affineWeyl$ satisfies $\lambda=w\cdot \Lambda$. This condition does not fix $w$ uniquely, but the RHS of equation \eqref{eq:klchar} depends on $w$ only modulo $\hat{\cW}_0$, the stabilizer of $\Lambda$ in $\affineWeyl$. The inequality $\le$ defining the sum is with regard to Bruhat ordering (see e.g. \cite{HumphreysCoxeter}). Finally, the coefficients $m_{w,w'}$ which replace the sign in the Weyl-Kac formula \eqref{eq:weyl-kac} are specializations of linear combinations of Kazhdan-Lusztig polynomials. Specifically,
\begin{equation*}
    m_{w,w'}=\left\{\begin{array}{cc}
        \tilde Q_{w,w'}(1) & \text{if } h^\vee_G-n>0 \\
        \tilde P_{w,w'}(1) & \text{if } h^\vee_G-n<0
    \end{array}\right.\,.
\end{equation*}
We will define only the polynomials $\tilde Q$ relevant for our computations at $h_G^\vee-n>0$ (see e.g. \cite{DeVos:1995an} for the definition of the polynomials $\tilde P$). They are given by

\begin{equation*}
    \tilde Q_{w,w'}=\sum_{z \in [w']}(-1)^{l(\bar w)+l(z)} Q_{\bar w,z} \,.
\end{equation*}
Here, $\bar w$ is a representative of maximal length of the class of $w$ in $\affineWeyl/\hat{\cW}_0$, the sum is over all $z$ in the class $[w']\in \affineWeyl/\hat{\cW}_0$ of $w'$, and $Q$ are the inverse Kazhdan-Lusztig polynomials. They can be computed recursively as follows:

\begin{enumerate}
    \item Initial data:
    \begin{align}
        Q_{w,w}&=1\,, \forall w \in \affineWeyl\,,\\ Q_{w_1,w_2}&=0 \,, \text{ if } w_1\nleq w_2\,.
    \end{align}
    \item Recursion step: For any simple reflection $s$
    \begin{align}
       Q_{w_1,w_2 s}&=Q_{w_1,w_2}\,,\quad \text{ if } w_1 s<w_1\,,  \\
       Q_{w_1,w_2 s}&=Q_{w_1 s,w_2}-q Q_{w_1,w_2}+q \sum_{\substack{w_1\leq w <w_2\\ws>w}}\slashed Q_{w_1,w}Q_{w,w_2}\,,\quad\text{ if } w_1 s>w_1\,,
    \end{align}
    where $q$ is the polynomial variable $Q$ depends on,
    and $\slashed Q_{w_1,w}$ is the highest order monomial in $Q_{w_1,w}$.
\end{enumerate}
Once all of this is in place, we can calculate the multiplicities $m_\lambda(\mu)$ of the weight $\mu$ in the highest weight representation $L_\lambda$ at negative level by comparing the coefficients of $e^\mu$ in equation \eqref{eq:klchar}. This yields
\begin{equation}
        m_\lambda(\mu)=\sum_{w\leq w'}m^M_{w'\cdot \Lambda}(\mu)\tilde Q_{w,w'}\,,
        \label{eq:klMult}
    \end{equation}
    where $m^M_{w'\cdot \Lambda}(\mu)$ denotes the multiplicity of $\mu$ in the Verma module of $w'\cdot \Lambda$. Note that the sum is finite because $m_{w'\cdot \Lambda}^M(\mu)$ is non-zero only if $\mu\leq w'\cdot \Lambda$.

    Based on the above, we implemented an algorithm to compute $\cha(L_\lambda)$ for $\lambda \in \dominant(\mathfrak{g})_{-n}$ in SageMath. The computation proceeds in two steps: first, we determine all weights up to a given grade which can appear in $L_\lambda$:
    \begin{enumerate}
        \item We consider the list of all weights of the form
    \begin{equation} \label{eq:weighttree}
        \lambda-k\alpha_i\,, \quad 0 <k \leq \lambda_i\,, \quad i=1,\dots,r
    \end{equation}
    with $\lambda_i$ the Dynkin labels of $\lambda$. For every weight $\omega$ in this list, we generate a new list following \eqref{eq:weighttree} with $\lambda$ replaced by $\omega$. We repeat this process until we find no further new weights. This gives all the weights at grade 0.
        \item We seed the above process at each new grade by subtracting the $0^{th}$ root $\alpha_0$ from all weights at the previous grade.
    \end{enumerate}
    To compute multiplicities, we decompose the weights at each grade into weight orbits of the finite Weyl group $\cW$, as multiplicities are invariant under the action of $\cW$. We choose the highest weight in each orbit. If it is not a null weight, we compute its multiplicity using Freudenthal's formula \eqref{eq:Freudenthal}. This is substantially faster than invoking \eqref{eq:klMult}, to which we resort in the case of null weights.

\section{The polynomials $p^\lambda_\omega$} \label{a:polynomials}

In this section, we list the polynomials $p^\lambda_\omega$ defined in equation \eqref{eq:poly_multiplying_omega} for numerous theories. The order to which we compute is encoded in the integers $M,m$ introduced in section \ref{sec:algorithm}. The polynomials are given in terms of $v$ and $y=q^{-\frac{1}{4\kappa}}$.

\subsection{$(C_r)_1$}
For $(C_r)_1$ the elliptic the functions $\xi$ can be written in terms of theta functions, as in equation \eqref{eq:xi_n=1_theta}. For the trivial and the vector representations, the $\xi$ functions only have $\theta_{3}=1+O(\sqrt{q})$ and $\theta_4=1+O(\sqrt{q})$ factors in the denominator therefore it is not necessary to make an expansion in $v$. We only need to expand in $q$ and we obtain results exact in $v$. We thus just truncate our expressions in $q$ and not in $v$. A conjecture for arbitrary $r$ is given in section \ref{sub:E-string results}. 

We use $0,v$ for the trivial and vector representation of $D_{8+2r}$ to ease notation. We only include the $0,v$ results because for these theories we know that applications $\mathcal{F}_\kappa$ give us the two spin contributions as explained in section \ref{sub:E-string results}.

For $r=3$ we computed the result to order $5+E_0$ in $q$. $\kappa=2$, so $y=q^{-\frac{1}{8}}$.
\begin{center}
     
    \scalebox{1}{$
\input{C3_1}$
}

\end{center}

For $r=4$ we computed the result to order $5+E_0$ in q. $\kappa=4$, so $y=q^{-\frac{1}{16}}$.
\begin{center}
     
\scalebox{1}{$
\input{C4_1}$
}

\end{center}

\subsection{$(B_3)_3$}
${(B_3)}_3$ was checked to order $M=10,m=13$. 
$\kappa=2$, so $y=q^{-\frac{1}{8}}$.

\begin{center}
     
\scalebox{1}{$
\input{B3_3}$
}

\end{center}
We note that $\mathcal F_\kappa(\xi_{(00)})=\xi_{(01)}$ and $\mathcal{F}_\kappa(\xi_{10})=\xi_{(10)}$.

\subsection{$(F_4)_3$}
${(F_4)}_3$ was checked to order $M=5,m=13$. 
$\kappa=6$, so $y=q^{-\frac{1}{24}}$.

\begin{center}
     
\scalebox{0.7}{$
\input{F4_3}$
}

\end{center}

\subsection{$(D_4)_4$}
$(D_4)_4$ was checked to order $M=5,m=8$. A conjecture for the general form is given in \ref{sub:matter-less}. $\kappa=2$, so $y=q^{-\frac{1}{8}}$.
\begin{center}
  
\scalebox{1}{$
\input{D4_4}$
}

\end{center}

We note $\mathcal F_\kappa(\xi_0)=-\xi_0$.

\subsection{${(B_4)}_4$}
${(B_4)}_4$ was checked to order $M=10,m=20$. A conjecture for the general form is given in section \ref{subsec:E7_7}.
$\kappa=3$, so $y=q^{-\frac{1}{12}}$.

\begin{center}
    
\scalebox{1}{$
\input{B4_4}$
}

\end{center}

We note that $\mathcal F_\kappa(\xi_{(0)})=-\xi_{(1)}$.

\subsection{${(D_5)}_4$}
${(D_5)}_4$ was checked to order $M=6,m=13$. 
$\kappa=4$, so $y=q^{-\frac{1}{16}}$.

\begin{center}
     
\scalebox{0.8}{$
\input{D5_4}$
}

\end{center}

We note that $\mathcal F_\kappa(\xi_{(00)})=-\xi_{(01)}$ and $\mathcal{F}_\kappa(\xi_{(10)})=\xi_{(10)}$.

\subsection{${(F_4)}_4$}
${(F_4)}_4$ was checked to order $M=9,m=19$. 
$\kappa=5$, so $y=q^{-\frac{1}{20}}$.

\begin{center}
     
\scalebox{0.55}{$
\input{F4_4}$
}

\end{center}

We note that $\mathcal F_\kappa(\xi_0)=\xi_3$ and $\mathcal{F}_\kappa(\xi_1)=\xi_2$.

\subsection{$(F_4)_5$}
$(F_4)_5$ was checked to order $M=10,m=20$. A conjecture for the general form is given in \ref{sub:matter-less}. $\kappa=4$, so $y=q^{-\frac{1}{16}}$.
\begin{center}
     
\scalebox{1}{$
\input{F4_5}$
}

\end{center}

We note $\mathcal F_\kappa(\xi_0)=\xi_0$.

\subsection{$(E_6)_5$} \label{aa:polynomialsE65}

$(E_6)_5$ was checked to order $M=7,m=16$. $\kappa=7$, so $y=q^{-\frac{1}{28}}$.

\begin{center}
     
\scalebox{0.65}{$
\input{E6_5}$
}

\end{center}

We note that $\mathcal F_\kappa(\xi_0)=\xi_3$ and $\mathcal{F}_\kappa(\xi_1+\xi_{-1})=\xi_{2}+\xi_{-2}$.

As announced in the text, $(E_6)_5$ is a theory for which the Dynkin symmetry akin ambiguity of the $U(1)$ flavor symmetry occurs. The flavor group here is $U(1)_6$. It has 6 integrable representations labeled by $l=-2,-1,0,1,2,3$. We cannot solve for $\xi_{\pm 1},\,\xi_{\pm 2}$ individually due to the $m\mapsto-m$ symmetry relating the characters $\hat \chi^{U(1)}_{\pm l}$. Above, we have displayed the Dynkin symmetric solution of equation \eqref{eq:tosolve} with regard to $E_6$, as the $(E_6)_5$ theory descends from a $E7$ theory with gauge group $E_7$.

\subsection{$(E_6)_6$}
$(E_6)_6$ was checked to order $M=9,m=18$. A conjecture for the general form is given in \ref{sub:matter-less}. $\kappa=6$, so $y=q^{-\frac{1}{24}}$.
\begin{center}

\scalebox{1}{$
\input{E6_6}$
}

\end{center}
We note that $\mathcal F_\kappa(\xi_0)=-\xi_0$.

\subsection{$(E_7)_7$}
$(E_7)_7$ was checked to order $M=9,m=20$. A conjecture for the general form and the action of $\mathcal F_\kappa$ are given in section \ref{subsec:E7_7}. $\kappa=11$, so $y=q^{-\frac{1}{44}}$.
\begin{center}
     
\scalebox{1}{$
\input{E7_7}$
}

\end{center}

\subsection{$(E_7)_8$}
$(E_7)_8$ was checked to order $M=10,m=18$. A conjecture for the general form is given in \ref{sub:matter-less}. $\kappa=10$, so $y=q^{-\frac{1}{40}}$.
\begin{center}
     
\scalebox{1}{$
\input{E7_8}$
}

\end{center}

We note that $\mathcal F_\kappa(\xi_0)=-\xi_0$.

\section{Higgsing trees} \label{a:higgsing_trees}
\label{sec:HiggsingTrees}

\input{higssing_trees}

\newpage

\bibliography{biblio}
    
\end{document}

%% file: C3_1.tex
\begin{array}{c|c}
 \xi _0 & \xi _v \\\hline
\begin{array}{cc}
 \hat{\chi }_{\text{(000)}} & 1 \\
 \hat{\chi }_{\text{(010)}} & 1 \\
 \hat{\chi }_{\text{(200)}} & v^2 y^4+\frac{y^4}{v^2} \\
 \hat{\chi }_{\text{(400)}} & v^4 y^{16}+\frac{y^{16}}{v^4} \\
 \hat{\chi }_{\text{(600)}} & v^6 y^{36}+\frac{y^{36}}{v^6} \\
 \hat{\chi }_{\text{(800)}} & v^8 y^{64}+\frac{y^{64}}{v^8} \\
 \hat{\chi }_{\text{(10,00)}} & v^{10} y^{100}+\frac{y^{100}}{v^{10}} \\
 \hat{\chi }_{\text{(12,00)}} & v^{12} y^{144}+\frac{y^{144}}{v^{12}} \\
\end{array}
& 
\begin{array}{cc}
 \hat{\chi }_{\text{(100)}} & -v y-\frac{y}{v} \\
 \hat{\chi }_{\text{(300)}} & -v^3 y^9-\frac{y^9}{v^3} \\
 \hat{\chi }_{\text{(500)}} & -v^5 y^{25}-\frac{y^{25}}{v^5} \\
 \hat{\chi }_{\text{(700)}} & -v^7 y^{49}-\frac{y^{49}}{v^7} \\
 \hat{\chi }_{\text{(900)}} & -v^9 y^{81}-\frac{y^{81}}{v^9} \\
 \hat{\chi }_{\text{(11,00)}} & -v^{11} y^{121}-\frac{y^{121}}{v^{11}} \\
\end{array}

\end{array}

%% file: C4_1.tex
\begin{array}{c|c}
 \xi _0 & \xi _v \\\hline
\begin{array}{cc}
 \hat{\chi }_{\text{(0000)}} & 1 \\
 \hat{\chi }_{\text{(0100)}} & 1 \\
 \hat{\chi }_{\text{(2000)}} & v^2 y^4+\frac{y^4}{v^2} \\
 \hat{\chi }_{\text{(4000)}} & v^4 y^{16}+\frac{y^{16}}{v^4} \\
 \hat{\chi }_{\text{(6000)}} & v^6 y^{36}+\frac{y^{36}}{v^6} \\
 \hat{\chi }_{\text{(8000)}} & v^8 y^{64}+\frac{y^{64}}{v^8} \\
 \hat{\chi }_{\text{(10,000)}} & v^{10} y^{100}+\frac{y^{100}}{v^{10}} \\
 \hat{\chi }_{\text{(12,000)}} & v^{12} y^{144}+\frac{y^{144}}{v^{12}} \\
\end{array}
&
\begin{array}{cc}
 \hat{\chi }_{\text{(1000)}} & -v y-\frac{y}{v} \\
 \hat{\chi }_{\text{(3000)}} & -v^3 y^9-\frac{y^9}{v^3} \\
 \hat{\chi }_{\text{(5000)}} & -v^5 y^{25}-\frac{y^{25}}{v^5} \\
 \hat{\chi }_{\text{(7000)}} & -v^7 y^{49}-\frac{y^{49}}{v^7} \\
 \hat{\chi }_{\text{(9000)}} & -v^9 y^{81}-\frac{y^{81}}{v^9} \\
 \hat{\chi }_{\text{(11,000)}} & -v^{11} y^{121}-\frac{y^{121}}{v^{11}} \\
\end{array}
\end{array}

%% file: B3_3.tex
\begin{array}{c|c|c}
 \xi _{\text{(00)}} & \xi _{\text{(01)}} & \xi _{\text{(10)}} \\\hline
 
\begin{array}{cc}
 \hat{\chi }_{\text{(000)}} & \frac{y^4}{v^2}-v^2 y^4 \\
 \hat{\chi }_{\text{(002)}} & v^6 y^{36}-v^2 y^4 \\
 \hat{\chi }_{\text{(010)}} & \frac{y^{16}}{v^4}-1 \\
 \hat{\chi }_{\text{(012)}} & v^8 y^{64} \\
 \hat{\chi }_{\text{(020)}} & \frac{y^{36}}{v^6}-\frac{y^4}{v^2} \\
 \hat{\chi }_{\text{(022)}} & v^{10} y^{100} \\
 \hat{\chi }_{\text{(030)}} & \frac{y^{64}}{v^8} \\
 \hat{\chi }_{\text{(032)}} & v^{12} y^{144} \\
 \hat{\chi }_{\text{(040)}} & \frac{y^{100}}{v^{10}} \\
 \hat{\chi }_{\text{(050)}} & \frac{y^{144}}{v^{12}} \\
 \hat{\chi }_{\text{(060)}} & \frac{y^{196}}{v^{14}} \\
 \hat{\chi }_{\text{(070)}} & \frac{y^{256}}{v^{16}} \\
 \hat{\chi }_{\text{(080)}} & \frac{y^{324}}{v^{18}} \\
 \hat{\chi }_{\text{(090)}} & \frac{y^{400}}{v^{20}} \\
 \hat{\chi }_{\text{(100)}} & v^4 y^{16}-1 \\
\end{array}
 & 
\begin{array}{cc}
 \hat{\chi }_{\text{(000)}} & v^4 y^{16}-1 \\
 \hat{\chi }_{\text{(002)}} & \frac{y^{16}}{v^4}-1 \\
 \hat{\chi }_{\text{(010)}} & v^6 y^{36}-v^2 y^4 \\
 \hat{\chi }_{\text{(012)}} & \frac{y^{36}}{v^6} \\
 \hat{\chi }_{\text{(020)}} & v^8 y^{64} \\
 \hat{\chi }_{\text{(022)}} & \frac{y^{64}}{v^8} \\
 \hat{\chi }_{\text{(030)}} & v^{10} y^{100} \\
 \hat{\chi }_{\text{(032)}} & \frac{y^{100}}{v^{10}} \\
 \hat{\chi }_{\text{(040)}} & v^{12} y^{144} \\
 \hat{\chi }_{\text{(042)}} & \frac{y^{144}}{v^{12}} \\
 \hat{\chi }_{\text{(052)}} & \frac{y^{196}}{v^{14}} \\
 \hat{\chi }_{\text{(062)}} & \frac{y^{256}}{v^{16}} \\
 \hat{\chi }_{\text{(072)}} & \frac{y^{324}}{v^{18}} \\
 \hat{\chi }_{\text{(100)}} & \frac{y^4}{v^2}-v^2 y^4 \\
\end{array}
 & 
\begin{array}{cc}
 \hat{\chi }_{\text{(001)}} & -v^5 y^{25}-\frac{y^9}{v^3}+2 v y \\
 \hat{\chi }_{\text{(011)}} & -v^7 y^{49}-\frac{y^{25}}{v^5}+v^3 y^9+\frac{y}{v} \\
 \hat{\chi }_{\text{(021)}} & -v^9 y^{81}-\frac{y^{49}}{v^7} \\
 \hat{\chi }_{\text{(031)}} & -v^{11} y^{121}-\frac{y^{81}}{v^9} \\
 \hat{\chi }_{\text{(041)}} & -v^{13} y^{169}-\frac{y^{121}}{v^{11}} \\
 \hat{\chi }_{\text{(051)}} & -\frac{y^{169}}{v^{13}} \\
 \hat{\chi }_{\text{(061)}} & -\frac{y^{225}}{v^{15}} \\
 \hat{\chi }_{\text{(071)}} & -\frac{y^{289}}{v^{17}} \\
 \hat{\chi }_{\text{(081)}} & -\frac{y^{361}}{v^{19}} \\
\end{array}
 \\
\end{array}

%% file: F4_3.tex
\begin{array}{c|c|c|c}
 \xi _{\text{(00)}} & \xi _{\text{(01)}} & \xi _{\text{(02)}} & \xi _{\text{(03)}} \\\hline
 
\begin{array}{cc}
 \hat{\chi }_{\text{(0000)}} & \frac{y^4}{v^2} \\
 \hat{\chi }_{\text{(0001)}} & v^2 y^4 \\
 \hat{\chi }_{\text{(0100)}} & \frac{y^4}{v^2}-v^{10} y^{100} \\
 \hat{\chi }_{\text{(1000)}} & \frac{y^{16}}{v^4}-v^8 y^{64} \\
 \hat{\chi }_{\text{(2000)}} & \frac{y^{36}}{v^6}-v^6 y^{36} \\
 \hat{\chi }_{\text{(3000)}} & \frac{y^{64}}{v^8} \\
 \hat{\chi }_{\text{(4000)}} & \frac{y^{100}}{v^{10}} \\
\end{array}
 & 
\begin{array}{cc}
 \hat{\chi }_{\text{(0000)}} & v^6 y^{36} \\
 \hat{\chi }_{\text{(0001)}} & v^6 y^{36} \\
 \hat{\chi }_{\text{(0002)}} & \frac{y^{16}}{v^4}-v^8 y^{64} \\
 \hat{\chi }_{\text{(0020)}} & v^{12} y^{144} \\
 \hat{\chi }_{\text{(1002)}} & \frac{y^{36}}{v^6} \\
 \hat{\chi }_{\text{(1100)}} & \frac{y^{36}}{v^6} \\
 \hat{\chi }_{\text{(2002)}} & \frac{y^{64}}{v^8} \\
\end{array}
 & 
\begin{array}{cc}
 \hat{\chi }_{\text{(0000)}} & 1 \\
 \hat{\chi }_{\text{(0001)}} & 1 \\
 \hat{\chi }_{\text{(0002)}} & v^{10} y^{100}-\frac{y^4}{v^2} \\
 \hat{\chi }_{\text{(0020)}} & \frac{y^{36}}{v^6} \\
 \hat{\chi }_{\text{(1002)}} & v^{12} y^{144} \\
\end{array}
 & 
\begin{array}{cc}
 \hat{\chi }_{\text{(0000)}} & v^8 y^{64} \\
 \hat{\chi }_{\text{(0001)}} & v^4 y^{16} \\
 \hat{\chi }_{\text{(0100)}} & -\frac{y^{16}}{v^4} \\
 \hat{\chi }_{\text{(1000)}} & v^{10} y^{100}-\frac{y^4}{v^2} \\
 \hat{\chi }_{\text{(2000)}} & v^{12} y^{144} \\
\end{array}
 \\\hline
 \xi _{\text{(10)}} & \xi _{\text{(11)}} & \xi _{\text{(12)}} & \xi _{\text{(20)}} \\\hline
 
\begin{array}{cc}
 \hat{\chi }_{\text{(0000)}} & -v^3 y^9 \\
 \hat{\chi }_{\text{(0001)}} & -\frac{y^9}{v^3} \\
 \hat{\chi }_{\text{(0010)}} & v^9 y^{81}-\frac{y^9}{v^3} \\
 \hat{\chi }_{\text{(0012)}} & -v^{11} y^{121} \\
 \hat{\chi }_{\text{(0110)}} & -v^{13} y^{169} \\
 \hat{\chi }_{\text{(1001)}} & v^7 y^{49}-\frac{y^{25}}{v^5} \\
 \hat{\chi }_{\text{(2001)}} & -\frac{y^{49}}{v^7}-\frac{y^{25}}{v^5} \\
 \hat{\chi }_{\text{(3001)}} & -\frac{y^{81}}{v^9} \\
\end{array}
 & 
\begin{array}{cc}
 \hat{\chi }_{\text{(0011)}} & -v^{11} y^{121}+v^7 y^{49}-\frac{y^{25}}{v^5}+\frac{y}{v} \\
 \hat{\chi }_{\text{(1011)}} & -v^{13} y^{169}-\frac{y^{49}}{v^7} \\
\end{array}
 & 
\begin{array}{cc}
 \hat{\chi }_{\text{(0000)}} & -v^3 y^9 \\
 \hat{\chi }_{\text{(0001)}} & -v^9 y^{81} \\
 \hat{\chi }_{\text{(0010)}} & \frac{y^9}{v^3}-v^9 y^{81} \\
 \hat{\chi }_{\text{(1001)}} & \frac{y}{v}-v^{11} y^{121} \\
 \hat{\chi }_{\text{(2001)}} & -v^{13} y^{169} \\
\end{array}
 & 
\begin{array}{cc}
 \hat{\chi }_{\text{(0000)}} & -v^4 y^{16} \\
 \hat{\chi }_{\text{(0001)}} & -v^8 y^{64} \\
 \hat{\chi }_{\text{(0003)}} & v^{10} y^{100}-\frac{y^4}{v^2} \\
 \hat{\chi }_{\text{(0010)}} & \frac{y^{16}}{v^4}-v^8 y^{64} \\
 \hat{\chi }_{\text{(0101)}} & v^{12} y^{144} \\
 \hat{\chi }_{\text{(1002)}} & \frac{y^{16}}{v^4} \\
 \hat{\chi }_{\text{(1010)}} & \frac{y^{36}}{v^6} \\
 \hat{\chi }_{\text{(2010)}} & \frac{y^{64}}{v^8} \\
\end{array}
 \\\hline
 &\xi _{\text{(21)}} & \xi _{\text{(30)}}&\\\hline
 &
\begin{array}{cc}
 \hat{\chi }_{\text{(0000)}} & -v^2 y^4 \\
 \hat{\chi }_{\text{(0001)}} & -\frac{y^4}{v^2} \\
 \hat{\chi }_{\text{(0003)}} & \frac{y^{16}}{v^4} \\
 \hat{\chi }_{\text{(0010)}} & v^{10} y^{100}-\frac{y^4}{v^2} \\
 \hat{\chi }_{\text{(0101)}} & \frac{y^{36}}{v^6} \\
 \hat{\chi }_{\text{(1010)}} & v^{12} y^{144} \\
\end{array}
 & 
\begin{array}{cc}
 \hat{\chi }_{\text{(0000)}} & v^7 y^{49}+\frac{y}{v} \\
 \hat{\chi }_{\text{(0001)}} & v^5 y^{25}+v y \\
 \hat{\chi }_{\text{(0100)}} & -v^{11} y^{121}+v^7 y^{49}-\frac{y^{25}}{v^5}+\frac{y}{v} \\
 \hat{\chi }_{\text{(1002)}} & -v^{11} y^{121}-\frac{y^{25}}{v^5} \\
 \hat{\chi }_{\text{(1100)}} & -v^{13} y^{169}-\frac{y^{49}}{v^7} \\
\end{array}&
 
\end{array}

%% file: D4_4.tex
\begin{array}{c}
 \xi _0 \\\hline
\begin{array}{cc}
 \hat{\chi }_{\text{(0000)}} & v^5 y^{25}+2 v^3 y^9-\frac{y^9}{v^3}-\frac{2 y}{v} \\
 \hat{\chi }_{\text{(0100)}} & v^7 y^{49}+2 v^5 y^{25}-\frac{y^{25}}{v^5}+v^3 y^9-\frac{2 y^9}{v^3}-\frac{y}{v} \\
 \hat{\chi }_{\text{(0200)}} & v^9 y^{81}+2 v^7 y^{49}-\frac{y^{49}}{v^7}+v^5 y^{25}-\frac{2 y^{25}}{v^5} \\
 \hat{\chi }_{\text{(0300)}} & 2 v^9 y^{81}-\frac{y^{81}}{v^9} \\
 \hat{\chi }_{\text{(0400)}} & -\frac{y^{121}}{v^{11}} \\
 \hat{\chi }_{\text{(0500)}} & -\frac{y^{169}}{v^{13}} \\
 \hat{\chi }_{\text{(0600)}} & -\frac{y^{225}}{v^{15}} \\
\end{array}
\end{array}

%% file: B4_4.tex
\begin{array}{c|c}
 \xi _{\text{(0)}} & \xi _{\text{(1)}} \\\hline
 
\begin{array}{cc}
 \hat{\chi }_{\text{(0000)}} & \frac{y^9}{v^3}-v^3 y^9 \\
 \hat{\chi }_{\text{(0100)}} & -v^5 y^{25}+\frac{y^{25}}{v^5}-v y+\frac{y}{v} \\
 \hat{\chi }_{\text{(0200)}} & -v^7 y^{49}+\frac{y^{49}}{v^7}+v y-\frac{y}{v} \\
 \hat{\chi }_{\text{(0300)}} & \frac{y^{81}}{v^9}-v^9 y^{81} \\
 \hat{\chi }_{\text{(0400)}} & \frac{y^{121}}{v^{11}} \\
 \hat{\chi }_{\text{(0500)}} & \frac{y^{169}}{v^{13}} \\
 \hat{\chi }_{\text{(0600)}} & \frac{y^{225}}{v^{15}} \\
 \hat{\chi }_{\text{(0700)}} & \frac{y^{289}}{v^{17}} \\
 \hat{\chi }_{\text{(0800)}} & \frac{y^{361}}{v^{19}} \\
 \hat{\chi }_{\text{(1000)}} & v^7 y^{49}+v^5 y^{25}-v y-\frac{y}{v} \\
 \hat{\chi }_{\text{(1100)}} & v^9 y^{81}-\frac{y^9}{v^3} \\
 \hat{\chi }_{\text{(1200)}} & v^{11} y^{121}-v^5 y^{25}-\frac{y^{25}}{v^5}+v y \\
 \hat{\chi }_{\text{(1300)}} & v^{13} y^{169}-\frac{y^{49}}{v^7} \\
 \hat{\chi }_{\text{(1400)}} & v^{15} y^{225} \\
 \hat{\chi }_{\text{(1500)}} & v^{17} y^{289} \\
 \hat{\chi }_{\text{(1600)}} & v^{19} y^{361} \\
 \hat{\chi }_{\text{(1700)}} & v^{21} y^{441} \\
 \hat{\chi }_{\text{(2000)}} & \frac{y^9}{v^3}-v^3 y^9 \\
\end{array}
 & 
\begin{array}{cc}
 \hat{\chi }_{\text{(0000)}} & 1-v^6 y^{36} \\
 \hat{\chi }_{\text{(0100)}} & -v^8 y^{64}-v^4 y^{16}+v^2 y^4+\frac{y^4}{v^2} \\
 \hat{\chi }_{\text{(0200)}} & -v^{10} y^{100}+v^4 y^{16}+\frac{y^{16}}{v^4}-v^2 y^4 \\
 \hat{\chi }_{\text{(0300)}} & \frac{y^{36}}{v^6}-v^{12} y^{144} \\
 \hat{\chi }_{\text{(0400)}} & -v^{14} y^{196} \\
 \hat{\chi }_{\text{(0500)}} & -v^{16} y^{256} \\
 \hat{\chi }_{\text{(0600)}} & -v^{18} y^{324} \\
 \hat{\chi }_{\text{(0700)}} & -v^{20} y^{400} \\
 \hat{\chi }_{\text{(1000)}} & v^4 y^{16}-\frac{y^{16}}{v^4}+v^2 y^4-\frac{y^4}{v^2} \\
 \hat{\chi }_{\text{(1100)}} & v^6 y^{36}-\frac{y^{36}}{v^6} \\
 \hat{\chi }_{\text{(1200)}} & v^8 y^{64}-\frac{y^{64}}{v^8}+\frac{y^4}{v^2} \\
 \hat{\chi }_{\text{(1300)}} & -\frac{y^{100}}{v^{10}} \\
 \hat{\chi }_{\text{(1400)}} & -\frac{y^{144}}{v^{12}} \\
 \hat{\chi }_{\text{(1500)}} & -\frac{y^{196}}{v^{14}} \\
 \hat{\chi }_{\text{(1600)}} & -\frac{y^{256}}{v^{16}} \\
 \hat{\chi }_{\text{(1700)}} & -\frac{y^{324}}{v^{18}} \\
 \hat{\chi }_{\text{(2000)}} & 1-v^6 y^{36} \\
\end{array}
 \\
\end{array}

%% file: D5_4.tex
\begin{array}{c|c|c}
 \xi _{\text{(00)}} & \xi _{\text{(01)}} & \xi _{\text{(10)}} \\\hline
 
\begin{array}{cc}
 \hat{\chi }_{\text{(00000)}} & v^5 y^{25}-2 v^3 y^9+\frac{y^9}{v^3} \\
 \hat{\chi }_{\text{(01000)}} & -2 v^5 y^{25}+\frac{y^{25}}{v^5}+v^3 y^9 \\
 \hat{\chi }_{\text{(02000)}} & \frac{y^{49}}{v^7}-v^9 y^{81} \\
 \hat{\chi }_{\text{(03000)}} & \frac{y^{81}}{v^9} \\
 \hat{\chi }_{\text{(04000)}} & \frac{y^{121}}{v^{11}} \\
 \hat{\chi }_{\text{(20000)}} & -v^9 y^{81}-v^7 y^{49}+v y+\frac{y}{v} \\
 \hat{\chi }_{\text{(21000)}} & \frac{y^{25}}{v^5}-v^{11} y^{121} \\
 \hat{\chi }_{\text{(22000)}} & -v^{13} y^{169} \\
\end{array}
 & 
\begin{array}{cc}
 \hat{\chi }_{\text{(00000)}} & -v^7 y^{49}+2 v y-\frac{y}{v} \\
 \hat{\chi }_{\text{(01000)}} & -v^9 y^{81}-v y+\frac{2 y}{v} \\
 \hat{\chi }_{\text{(02000)}} & \frac{y^{25}}{v^5}-v^{11} y^{121} \\
 \hat{\chi }_{\text{(03000)}} & -v^{13} y^{169} \\
 \hat{\chi }_{\text{(20000)}} & -v^5 y^{25}+\frac{y^{25}}{v^5}-v^3 y^9+\frac{y^9}{v^3} \\
 \hat{\chi }_{\text{(21000)}} & \frac{y^{49}}{v^7} \\
 \hat{\chi }_{\text{(22000)}} & \frac{y^{81}}{v^9} \\
\end{array}
 & 
\begin{array}{cc}
 \hat{\chi }_{\text{(10000)}} & v^8 y^{64}+v^4 y^{16}-\frac{y^{16}}{v^4}-1 \\
 \hat{\chi }_{\text{(11000)}} & v^{10} y^{100}+2 v^6 y^{36}-\frac{y^{36}}{v^6}-\frac{2 y^4}{v^2} \\
 \hat{\chi }_{\text{(12000)}} & v^{12} y^{144}-\frac{y^{64}}{v^8} \\
 \hat{\chi }_{\text{(13000)}} & v^{14} y^{196}-\frac{y^{100}}{v^{10}} \\
 \hat{\chi }_{\text{(30000)}} & v^8 y^{64}-\frac{y^{16}}{v^4} \\
\end{array}

\end{array}

%% file: F4_4.tex
\begin{array}{c|c|c|c}
 \xi _0 & \xi _1 & \xi _2 & \xi _3 \\\hline
 
\begin{array}{cc}
 \hat{\chi }_{\text{(0000)}} & \frac{y^9}{v^3}-v^7 y^{49} \\
 \hat{\chi }_{\text{(0002)}} & v^9 y^{81}-\frac{y}{v} \\
 \hat{\chi }_{\text{(0004)}} & \frac{y^9}{v^3}-v^7 y^{49} \\
 \hat{\chi }_{\text{(0100)}} & v^{11} y^{121}+v^9 y^{81}-v y-\frac{y}{v} \\
 \hat{\chi }_{\text{(1000)}} & \frac{y^{25}}{v^5}-v^5 y^{25} \\
 \hat{\chi }_{\text{(1100)}} & v^{13} y^{169}+\frac{y^{49}}{v^7}-2 v^3 y^9 \\
 \hat{\chi }_{\text{(2000)}} & -v^7 y^{49}+\frac{y^{49}}{v^7}-v^3 y^9+\frac{y^9}{v^3} \\
 \hat{\chi }_{\text{(2100)}} & v^{15} y^{225} \\
 \hat{\chi }_{\text{(3000)}} & v^{11} y^{121}+\frac{y^{81}}{v^9}-2 v y \\
 \hat{\chi }_{\text{(3100)}} & v^{17} y^{289} \\
 \hat{\chi }_{\text{(4000)}} & \frac{y^{121}}{v^{11}}+\frac{y^{81}}{v^9} \\
 \hat{\chi }_{\text{(4100)}} & v^{19} y^{361} \\
 \hat{\chi }_{\text{(5000)}} & v^{17} y^{289}+\frac{y^{169}}{v^{13}} \\
 \hat{\chi }_{\text{(6000)}} & \frac{y^{225}}{v^{15}} \\
 \hat{\chi }_{\text{(7000)}} & \frac{y^{289}}{v^{17}} \\
\end{array}
 & 
\begin{array}{cc}
 \hat{\chi }_{\text{(0001)}} & v^6 y^{36}-\frac{y^{16}}{v^4} \\
 \hat{\chi }_{\text{(0002)}} & \frac{y^4}{v^2}-v^8 y^{64} \\
 \hat{\chi }_{\text{(0003)}} & v^6 y^{36}-\frac{y^{16}}{v^4} \\
 \hat{\chi }_{\text{(0010)}} & 1-v^{10} y^{100} \\
 \hat{\chi }_{\text{(1001)}} & v^6 y^{36}-\frac{y^{36}}{v^6}+v^4 y^{16}-\frac{y^{16}}{v^4} \\
 \hat{\chi }_{\text{(1010)}} & -v^{12} y^{144}-v^8 y^{64}+v^2 y^4+\frac{y^4}{v^2} \\
 \hat{\chi }_{\text{(2001)}} & -v^{12} y^{144}-\frac{y^{64}}{v^8}+2 v^2 y^4 \\
 \hat{\chi }_{\text{(2010)}} & -v^{14} y^{196}-\frac{y^{36}}{v^6} \\
 \hat{\chi }_{\text{(3001)}} & -\frac{y^{100}}{v^{10}} \\
 \hat{\chi }_{\text{(3010)}} & -v^{16} y^{256}-v^{14} y^{196} \\
 \hat{\chi }_{\text{(4001)}} & -\frac{y^{144}}{v^{12}} \\
 \hat{\chi }_{\text{(4010)}} & -v^{18} y^{324}-\frac{y^{144}}{v^{12}} \\
 \hat{\chi }_{\text{(5001)}} & -\frac{y^{196}}{v^{14}} \\
 \hat{\chi }_{\text{(5010)}} & -v^{20} y^{400} \\
 \hat{\chi }_{\text{(6001)}} & -\frac{y^{256}}{v^{16}} \\
\end{array}
 & 
\begin{array}{cc}
 \hat{\chi }_{\text{(0001)}} & v^9 y^{81}-\frac{y}{v} \\
 \hat{\chi }_{\text{(0002)}} & \frac{y^9}{v^3}-v^7 y^{49} \\
 \hat{\chi }_{\text{(0003)}} & v^9 y^{81}-\frac{y}{v} \\
 \hat{\chi }_{\text{(0010)}} & \frac{y^{25}}{v^5}-v^5 y^{25} \\
 \hat{\chi }_{\text{(1001)}} & v^{11} y^{121}+v^9 y^{81}-v y-\frac{y}{v} \\
 \hat{\chi }_{\text{(1010)}} & -v^7 y^{49}+\frac{y^{49}}{v^7}-v^3 y^9+\frac{y^9}{v^3} \\
 \hat{\chi }_{\text{(2001)}} & v^{13} y^{169}+\frac{y^{49}}{v^7}-2 v^3 y^9 \\
 \hat{\chi }_{\text{(2010)}} & v^{11} y^{121}+\frac{y^{81}}{v^9} \\
 \hat{\chi }_{\text{(3001)}} & v^{15} y^{225} \\
 \hat{\chi }_{\text{(3010)}} & \frac{y^{121}}{v^{11}}+\frac{y^{81}}{v^9} \\
 \hat{\chi }_{\text{(4001)}} & v^{17} y^{289} \\
 \hat{\chi }_{\text{(4010)}} & \frac{y^{169}}{v^{13}} \\
 \hat{\chi }_{\text{(5001)}} & v^{19} y^{361} \\
 \hat{\chi }_{\text{(5010)}} & \frac{y^{225}}{v^{15}} \\
\end{array}
 & 
\begin{array}{cc}
 \hat{\chi }_{\text{(0000)}} & \frac{y^4}{v^2}-v^8 y^{64} \\
 \hat{\chi }_{\text{(0002)}} & v^6 y^{36}-\frac{y^{16}}{v^4} \\
 \hat{\chi }_{\text{(0004)}} & \frac{y^4}{v^2}-v^8 y^{64} \\
 \hat{\chi }_{\text{(0100)}} & v^6 y^{36}-\frac{y^{36}}{v^6}+v^4 y^{16}-\frac{y^{16}}{v^4} \\
 \hat{\chi }_{\text{(1000)}} & 1-v^{10} y^{100} \\
 \hat{\chi }_{\text{(1100)}} & -v^{12} y^{144}-\frac{y^{64}}{v^8}+2 v^2 y^4 \\
 \hat{\chi }_{\text{(2000)}} & -v^{12} y^{144}-v^8 y^{64}+v^2 y^4+\frac{y^4}{v^2} \\
 \hat{\chi }_{\text{(2100)}} & -\frac{y^{100}}{v^{10}} \\
 \hat{\chi }_{\text{(3000)}} & -v^{14} y^{196}-\frac{y^{36}}{v^6} \\
 \hat{\chi }_{\text{(3100)}} & -\frac{y^{144}}{v^{12}} \\
 \hat{\chi }_{\text{(4000)}} & -v^{16} y^{256}-v^{14} y^{196} \\
 \hat{\chi }_{\text{(4100)}} & -\frac{y^{196}}{v^{14}} \\
 \hat{\chi }_{\text{(5000)}} & -v^{18} y^{324}-\frac{y^{144}}{v^{12}} \\
 \hat{\chi }_{\text{(6000)}} & -v^{20} y^{400} \\
\end{array}
 \\
\end{array}

%% file: F4_5.tex
\begin{array}{c}
 \xi _0 \\\hline
\begin{array}{cc}
 \hat{\chi }_{\text{(0000)}} & v^8 y^{64}-v^4 y^{16}+\frac{y^{16}}{v^4}-1 \\
 \hat{\chi }_{\text{(1000)}} & v^{10} y^{100}+\frac{y^{36}}{v^6}-2 v^2 y^4 \\
 \hat{\chi }_{\text{(2000)}} & v^{12} y^{144}+\frac{y^{64}}{v^8}-v^4 y^{16}-1 \\
 \hat{\chi }_{\text{(3000)}} & v^{14} y^{196}+\frac{y^{100}}{v^{10}} \\
 \hat{\chi }_{\text{(4000)}} & v^{16} y^{256}+\frac{y^{144}}{v^{12}} \\
 \hat{\chi }_{\text{(5000)}} & v^{18} y^{324}+\frac{y^{196}}{v^{14}} \\
 \hat{\chi }_{\text{(6000)}} & v^{20} y^{400}+\frac{y^{256}}{v^{16}} \\
 \hat{\chi }_{\text{(7000)}} & v^{22} y^{484}+\frac{y^{324}}{v^{18}} \\
\end{array}
\end{array}

%% file: E6_5.tex
\begin{array}{c|c|c|c}
 \xi _0 & \xi _{-1}+\xi_1 & \xi_{-2}+\xi _2 & \xi _3 \\\hline
 
\begin{array}{cc}
 \hat{\chi }_{\text{(000000)}} & v^{10} y^{100}-2 v^4 y^{16}+\frac{y^{16}}{v^4} \\
 \hat{\chi }_{\text{(000001)}} & v^8 y^{64}-2 v^6 y^{36}+\frac{y^{36}}{v^6} \\
 \hat{\chi }_{\text{(000002)}} & -2 v^8 y^{64}+\frac{y^{64}}{v^8}+v^6 y^{36} \\
 \hat{\chi }_{\text{(000003)}} & \frac{y^{100}}{v^{10}} \\
 \hat{\chi }_{\text{(000004)}} & \frac{y^{144}}{v^{12}} \\
 \hat{\chi }_{\text{(001000)}} & 1-v^{14} y^{196} \\
 \hat{\chi }_{\text{(001001)}} & -v^{16} y^{256} \\
 \hat{\chi }_{\text{(001002)}} & -v^{18} y^{324} \\
 \hat{\chi }_{\text{(100010)}} & -v^{12} y^{144}+2 v^2 y^4-\frac{y^4}{v^2} \\
\end{array}
 & 
\begin{array}{cc}
 \hat{\chi }_{\text{(000010)}} & -v^9 y^{81}+2 v^5 y^{25}-\frac{y^{25}}{v^5} \\
 \hat{\chi }_{\text{(000011)}} & v^7 y^{49}-\frac{y^{49}}{v^7} \\
 \hat{\chi }_{\text{(000012)}} & -\frac{y^{81}}{v^9} \\
 \hat{\chi }_{\text{(000013)}} & -\frac{y^{121}}{v^{11}} \\
 \hat{\chi }_{\text{(000020)}} & v^{11} y^{121}-2 v^3 y^9+\frac{y^9}{v^3} \\
 \hat{\chi }_{\text{(000100)}} & v^{13} y^{169}-2 v y+\frac{y}{v} \\
 \hat{\chi }_{\text{(000101)}} & v^{15} y^{225} \\
 \hat{\chi }_{\text{(000102)}} & v^{17} y^{289} \\
 \hat{\chi }_{\text{(010000)}} & v^{13} y^{169}-2 v y+\frac{y}{v} \\
 \hat{\chi }_{\text{(010001)}} & v^{15} y^{225} \\
 \hat{\chi }_{\text{(010002)}} & v^{17} y^{289} \\
 \hat{\chi }_{\text{(100000)}} & -v^9 y^{81}+2 v^5 y^{25}-\frac{y^{25}}{v^5} \\
 \hat{\chi }_{\text{(100001)}} & v^7 y^{49}-\frac{y^{49}}{v^7} \\
 \hat{\chi }_{\text{(100002)}} & -\frac{y^{81}}{v^9} \\
 \hat{\chi }_{\text{(100003)}} & -\frac{y^{121}}{v^{11}} \\
 \hat{\chi }_{\text{(200000)}} & v^{11} y^{121}-2 v^3 y^9+\frac{y^9}{v^3} \\
\end{array}
 & 
\begin{array}{cc}
 \hat{\chi }_{\text{(000010)}} & -v^{12} y^{144}+2 v^2 y^4-\frac{y^4}{v^2} \\
 \hat{\chi }_{\text{(000011)}} & 1-v^{14} y^{196} \\
 \hat{\chi }_{\text{(000012)}} & -v^{16} y^{256} \\
 \hat{\chi }_{\text{(000013)}} & -v^{18} y^{324} \\
 \hat{\chi }_{\text{(000020)}} & v^{10} y^{100}-2 v^4 y^{16}+\frac{y^{16}}{v^4} \\
 \hat{\chi }_{\text{(000100)}} & v^8 y^{64}-2 v^6 y^{36}+\frac{y^{36}}{v^6} \\
 \hat{\chi }_{\text{(000101)}} & \frac{y^{64}}{v^8} \\
 \hat{\chi }_{\text{(000102)}} & \frac{y^{100}}{v^{10}} \\
 \hat{\chi }_{\text{(010000)}} & v^8 y^{64}-2 v^6 y^{36}+\frac{y^{36}}{v^6} \\
 \hat{\chi }_{\text{(010001)}} & \frac{y^{64}}{v^8} \\
 \hat{\chi }_{\text{(010002)}} & \frac{y^{100}}{v^{10}} \\
 \hat{\chi }_{\text{(100000)}} & -v^{12} y^{144}+2 v^2 y^4-\frac{y^4}{v^2} \\
 \hat{\chi }_{\text{(100001)}} & 1-v^{14} y^{196} \\
 \hat{\chi }_{\text{(100002)}} & -v^{16} y^{256} \\
 \hat{\chi }_{\text{(100003)}} & -v^{18} y^{324} \\
 \hat{\chi }_{\text{(200000)}} & v^{10} y^{100}-2 v^4 y^{16}+\frac{y^{16}}{v^4} \\
\end{array}
 & 
\begin{array}{cc}
 \hat{\chi }_{\text{(000000)}} & v^{11} y^{121}-2 v^3 y^9+\frac{y^9}{v^3} \\
 \hat{\chi }_{\text{(000001)}} & v^{13} y^{169}-2 v y+\frac{y}{v} \\
 \hat{\chi }_{\text{(000002)}} & v^{15} y^{225} \\
 \hat{\chi }_{\text{(000003)}} & v^{17} y^{289} \\
 \hat{\chi }_{\text{(001000)}} & -\frac{y^{49}}{v^7} \\
 \hat{\chi }_{\text{(001001)}} & -\frac{y^{81}}{v^9} \\
 \hat{\chi }_{\text{(100010)}} & -v^9 y^{81}+2 v^5 y^{25}-\frac{y^{25}}{v^5} \\
\end{array}
 \\
\end{array}

%% file: E6_6.tex
\begin{array}{c}
 \xi _0 \\\hline
\begin{array}{cc}
 \hat{\chi }_{\text{(000000)}} & -v^{11} y^{121}+v^7 y^{49}-2 v^5 y^{25}+\frac{y^{25}}{v^5}+2 v y-\frac{y}{v} \\
 \hat{\chi }_{\text{(000001)}} & -v^{13} y^{169}-2 v^7 y^{49}+\frac{y^{49}}{v^7}+v^5 y^{25}-v y+\frac{2 y}{v} \\
 \hat{\chi }_{\text{(000002)}} & -v^{15} y^{225}-2 v^9 y^{81}+\frac{y^{81}}{v^9}+\frac{2 y^9}{v^3} \\
 \hat{\chi }_{\text{(000003)}} & \frac{y^{121}}{v^{11}}-v^{17} y^{289} \\
 \hat{\chi }_{\text{(000004)}} & \frac{y^{169}}{v^{13}}-v^{19} y^{361} \\
 \hat{\chi }_{\text{(000005)}} & \frac{y^{225}}{v^{15}}-v^{21} y^{441} \\
\end{array}
\end{array}

%% file: E7_7.tex
\begin{array}{c}
 \xi _0 \\\hline
\begin{array}{cc}
 \hat{\chi }_{\text{(0000000)}} & v^{16} y^{256}-2 v^6 y^{36}+\frac{y^{36}}{v^6} \\
 \hat{\chi }_{\text{(0000001)}} & v^{19} y^{361}-2 v^3 y^9+\frac{y^9}{v^3} \\
 \hat{\chi }_{\text{(0000010)}} & -v^{15} y^{225}+2 v^7 y^{49}-\frac{y^{49}}{v^7} \\
 \hat{\chi }_{\text{(0000100)}} & -v^{18} y^{324}+2 v^4 y^{16}-\frac{y^{16}}{v^4} \\
 \hat{\chi }_{\text{(0001000)}} & v^{17} y^{289}-2 v^5 y^{25}+\frac{y^{25}}{v^5} \\
 \hat{\chi }_{\text{(0100000)}} & -v^{20} y^{400}+2 v^2 y^4-\frac{y^4}{v^2} \\
 \hat{\chi }_{\text{(1000000)}} & v^{14} y^{196}-2 v^8 y^{64}+\frac{y^{64}}{v^8} \\
 \hat{\chi }_{\text{(1000001)}} & v^{21} y^{441} \\
 \hat{\chi }_{\text{(1000010)}} & -v^{13} y^{169}+2 v^9 y^{81}-\frac{y^{81}}{v^9} \\
 \hat{\chi }_{\text{(1100000)}} & -v^{22} y^{484} \\
 \hat{\chi }_{\text{(2000000)}} & v^{12} y^{144}-2 v^{10} y^{100}+\frac{y^{100}}{v^{10}} \\
 \hat{\chi }_{\text{(2000001)}} & v^{23} y^{529} \\
 \hat{\chi }_{\text{(2000010)}} & -\frac{y^{121}}{v^{11}} \\
 \hat{\chi }_{\text{(2100000)}} & -v^{24} y^{576} \\
 \hat{\chi }_{\text{(3000000)}} & \frac{y^{144}}{v^{12}} \\
 \hat{\chi }_{\text{(3000010)}} & -\frac{y^{169}}{v^{13}} \\
 \hat{\chi }_{\text{(4000000)}} & \frac{y^{196}}{v^{14}} \\
\end{array} 
\end{array}

%% file: E7_8.tex
\begin{array}{c}
 \xi _0 \\\hline

\begin{array}{cc}
 \hat{\chi }_{\text{(0000000)}} & -v^{17} y^{289}+v^{13} y^{169}-2 v^7 y^{49}+\frac{y^{49}}{v^7}+2 v^3 y^9-\frac{y^9}{v^3} \\
 \hat{\chi }_{\text{(1000000)}} & -v^{19} y^{361}+v^{11} y^{121}-2 v^9 y^{81}+\frac{y^{81}}{v^9}+2 v y-\frac{y}{v} \\
 \hat{\chi }_{\text{(2000000)}} & \frac{y^{121}}{v^{11}}-v^{21} y^{441} \\
 \hat{\chi }_{\text{(3000000)}} & \frac{y^{169}}{v^{13}}-v^{23} y^{529} \\
 \hat{\chi }_{\text{(4000000)}} & \frac{y^{225}}{v^{15}} \\
\end{array}

\end{array}

%% file: higssing_trees.tex
We can organize the theories considered in this paper in trees where each line links theories associated by Higgsing. This gives rise to the Higgsing trees we present below, one for each base Hirzebruch surface $\mathbb F_n$ with $n=1,\dots,8,12$. For each theory, we give the gauge group and the flavor group with the level of the corresponding current. We write the rank of the gauge groups beside the class and not as a sub index to lighten notation; for instance $C3_{-1}$ means the gauge group is $C_3$ and the corresponding current is at level $-1$.

Each theory is also assigned a color: In {\color{lime}lime }, we give the theories for which the constants $c$ in the affine ansatz \eqref{eq:affineAnsatzIntro} were computed in \cite{DelZotto:2018tcj}. In {\color{green} green}, we give the theories for which we have computed them or have a conjectural form. In {\color{blue} blue}, we give the theories for which our methods should give complete answers. In {\color{cyan} cyan}, we give the theories for which our methods can be used but the results obtained would still have an ambiguity due to the Dynkin symmetry as explained in section \ref{ss:strategy}. Finally, in {\color{red} red}, we give the theories for which $h^\vee_G-n\leq 0 $  so the ansatz \eqref{eq:affineAnsatzIntro} cannot be used \footnote{There are just a handful of theories for which this happens and alternative expressions for their elliptic genera where given in \cite{DelZotto:2018tcj}.}, theories for which we do not know the flavor group, or theories for which we do not expect Dynkin symmetry. The structure of the trees is reproduced form \cite{DelZotto:2018tcj}. 

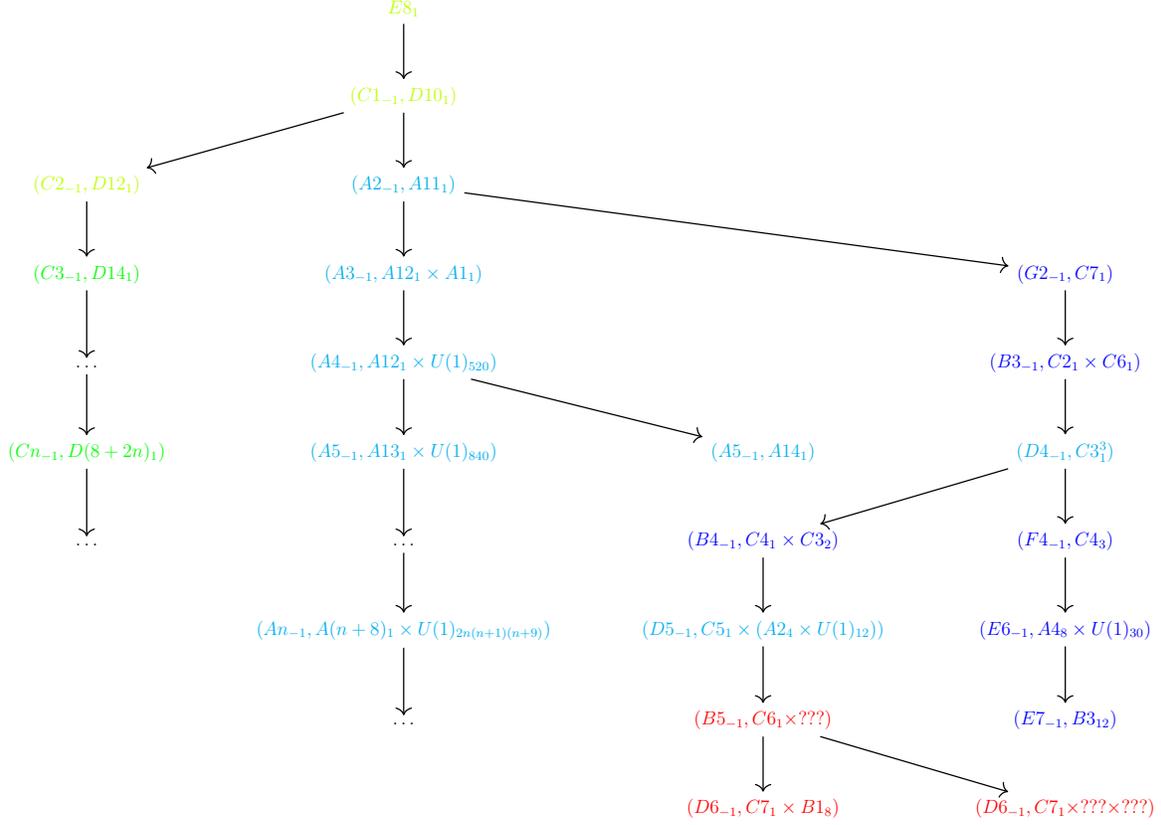
\begin{figure}
    \centering
\[\begin{tikzcd}[scale cd =0.6]
	& \textcolor{lime}{E8_1} \\
	& \textcolor{lime}{(C1_{-1},D10_{1})} \\
	\textcolor{lime}{(C2_{-1},D12_1)} & \textcolor{cyan}{(A2_{-1},A11_1)} \\
	\textcolor{green}{(C3_{-1},D14_1)} & \textcolor{cyan}{(A3_{-1},A12_1\times A1_1)} && \textcolor{blue}{(G2_{-1},C7_1)} \\
	\dots & \textcolor{cyan}{(A4_{-1},A12_1\times U(1)_{520})} && \textcolor{blue}{(B3_{-1},C2_1\times C6_1)} \\
	\textcolor{green}{(Cn_{-1},D(8+2n)_1)} & \textcolor{cyan}{(A5_{-1},A13_1\times U(1)_{840})} & \textcolor{cyan}{(A5_{-1},A14_1)} & \textcolor{cyan}{(D4_{-1},C3_1^3)} \\
	\dots & \dots & \textcolor{blue}{(B4_{-1},C4_1\times C3_2)} & \textcolor{blue}{(F4_{-1},C4_3)} \\
	& \textcolor{cyan}{(An_{-1},A(n+8)_1\times U(1)_{2n(n+1)(n+9)})} & \textcolor{cyan}{(D5_{-1},C5_1\times(A2_4\times U(1)_{12}))} & \textcolor{blue}{(E6_{-1},A4_8\times U(1)_{30})} \\
	& \dots & \textcolor{red}{(B5_{-1},C6_1\times???)} & \textcolor{blue}{(E7_{-1},B3_{12})} \\
	&& \textcolor{red}{(D6_{-1},C7_1\times B1_8)} & \textcolor{red}{(D6_{-1},C7_1\times???\times???)}
	\arrow[from=1-2, to=2-2]
	\arrow[from=2-2, to=3-1]
	\arrow[from=3-1, to=4-1]
	\arrow[from=4-1, to=5-1]
	\arrow[from=2-2, to=3-2]
	\arrow[from=3-2, to=4-2]
	\arrow[from=4-2, to=5-2]
	\arrow[from=5-1, to=6-1]
	\arrow[from=6-1, to=7-1]
	\arrow[from=5-2, to=6-2]
	\arrow[from=5-2, to=6-3]
	\arrow[from=6-2, to=7-2]
	\arrow[from=7-2, to=8-2]
	\arrow[from=8-2, to=9-2]
	\arrow[from=3-2, to=4-4]
	\arrow[from=4-4, to=5-4]
	\arrow[from=5-4, to=6-4]
	\arrow[from=6-4, to=7-4]
	\arrow[from=6-4, to=7-3]
	\arrow[from=7-3, to=8-3]
	\arrow[from=8-3, to=9-3]
	\arrow[from=9-3, to=10-3]
	\arrow[from=9-3, to=10-4]
	\arrow[from=7-4, to=8-4]
	\arrow[from=8-4, to=9-4]
\end{tikzcd}\]
    \caption{Higgsing tree of $\mathbb F_1$, the E-string. The root is the E-string with its $E_8$ flavor group.}
    \label{fig:F1higgsingtree}
\end{figure}

\begin{figure}
    \centering
    \begin{subfigure}[b]{0.6\textwidth}
    \centering
\[\begin{tikzcd}[scale cd =0.4]
&\textcolor{red}{A_1-(0,2)}\\
	& \textcolor{red}{(A1_{-2},B3_1\times Ising)} \\
	&  \textcolor{red}{(A1_{-2},D4_1)} \\
	& \textcolor{cyan}{(A2_{-2},A5_1)} \\
	\textcolor{cyan}{(A3_{-2},A7_{1})} & \textcolor{blue}{(G2_{-2},C4_1)} \\
	\dots & \textcolor{blue}{(B3_{-2},C1_1\times C4_1)} \\
	\textcolor{cyan}{(An_{-2},A(2n+1)_1)} & \textcolor{cyan}{(D4_{-2},C2_1^3)} \\
	\dots& \textcolor{blue}{(F4_{-2},C3_3)} & \textcolor{blue}{(B4_{-2},C2_1^3)} \\
	& \textcolor{cyan}{(E6_{-2},A3_6\times U(1)_{24})} & \textcolor{cyan}{(D5_{-2},C4_1\times A1_4\times U(1)_8)} \\
	& \textcolor{blue}{(E7_{-2},A3_{12})} & \textcolor{red}{(B5_{-2},C5_1\times???)} \\
	& \textcolor{red}{(D6_{-2},C6_1\times(Ising)\times(Ising))} & \textcolor{cyan}{(D6_{-2},C6_1\times A1_8)} \\
	&& \textcolor{blue}{(B6_{-2},C7_1)}
    \arrow[from=2-2, to=3-2]
	\arrow[from=3-2, to=4-2]
	\arrow[from=4-2, to=5-1]
	\arrow[from=5-1, to=6-1]
	\arrow[from=6-1, to=7-1]
	\arrow[from=7-1, to=8-1]
	\arrow[from=4-2, to=5-2]
	\arrow[from=5-2, to=6-2]
	\arrow[from=6-2, to=7-2]
	\arrow[from=7-2, to=8-2]
	\arrow[from=8-2, to=9-2]
	\arrow[from=9-2, to=10-2]
	\arrow[from=7-2, to=8-3]
	\arrow[from=8-3, to=9-3]
	\arrow[from=9-3, to=10-3]
	\arrow[from=10-3, to=11-3]
	\arrow[from=10-3, to=11-2]
	\arrow[from=11-3, to=12-3]
	\arrow[from=1-2, to=2-2]
\end{tikzcd}\]
    \caption{Higgsing tree of $\mathbb F_2$. The M-string Higgsing tree. $A_1-(0,2)$ stands for the 6D theory with $(0,2)$ supersymmetry.}
    \label{fig:F2higgsingtree}
    \end{subfigure}
    \hfill
    \begin{subfigure}[b]{0.35\textwidth}
    \centering
    \[\begin{tikzcd}[scale cd =0.4]
	\textcolor{red}{A2_{-3}} \\
	\textcolor{lime}{(G2_{-3},C1_1)} \\
	\textcolor{green}{(B3_{-3},C2_1)} \\
	\textcolor{cyan}{(D4_{-3},C1_1^3)} \\
	\textcolor{green}{(F4_{-3},C2_3)} & \textcolor{blue}{(B4_{-3},C2_1\times C1_2)} \\
	\textcolor{cyan}{(E6_{-3},A2_6\times U(1)_{18})} & \textcolor{red}{(D5_{-3},C3_2\times SU(1)_4\times U(1)_4)} \\
	\textcolor{blue}{(E7,C2_{12})} & \textcolor{blue}{(B5_{-3},C4_1\times (Ising))} \\
	& \textcolor{red}{(D6_{-3},C5_1)}
	\arrow[from=1-1, to=2-1]
	\arrow[from=2-1, to=3-1]
	\arrow[from=3-1, to=4-1]
	\arrow[from=4-1, to=5-1]
	\arrow[from=5-1, to=6-1]
	\arrow[from=6-1, to=7-1]
	\arrow[from=4-1, to=5-2]
	\arrow[from=5-2, to=6-2]
	\arrow[from=6-2, to=7-2]
	\arrow[from=7-2, to=8-2]
\end{tikzcd}\]
     \caption{Higgsing tree of $\mathbb F_3$.}
    \label{fig:F3higgsingtree}
\end{subfigure}
\hfill
    \begin{subfigure}{0.45\textwidth}
\[\begin{tikzcd}[scale cd =0.5]
	\textcolor{green}{D4_{-4}} \\
	\textcolor{green}{(B4_{-4},C1_1)} & \textcolor{green}{(F4_{-4},C1_3)} \\
	\textcolor{green}{(D5_{-4},C2_1)} & \textcolor{cyan}{(E6_{-4},A1_6\times U(1)_{12})} \\
	\textcolor{green}{(B5_{-4},C3_1)} & \textcolor{blue}{(E7_{-4},SO(4)_{12})} \\
	\dots \\
	\textcolor{blue}{(SO(N)_{-4},C(N-8)_1)} \\
	\dots
	\arrow[from=1-1, to=2-1]
	\arrow[from=2-1, to=3-1]
	\arrow[from=3-1, to=4-1]
	\arrow[from=1-1, to=2-2]
	\arrow[from=2-2, to=3-2]
	\arrow[from=4-1, to=5-1]
	\arrow[from=3-2, to=4-2]
	\arrow[from=5-1, to=6-1]
	\arrow[from=6-1, to=7-1]
\end{tikzcd}\]
     \caption{Higgsing tree of $\mathbb F_4$.}
    \label{fig:F4higgsingtree}
    \end{subfigure}
    \begin{subfigure}{0.45\textwidth}
    \centering
\[\begin{tikzcd}[scale cd =0.5]
	\textcolor{blue}{E8_{12}} \\
	\textcolor{green}{E7_8} \\
	\textcolor{blue}{E7_7} \\
	\textcolor{green}{E6_6} & \textcolor{cyan}{(E7_6,U(1)_{12})} & {} \\
	\textcolor{green}{F4_5} & \textcolor{green}{(E6_5,U(1)_6)} & \textcolor{blue}{(E7_5,SO(3)_{12})}
	\arrow[from=4-1, to=4-2]
	\arrow[from=5-2, to=5-3]
	\arrow[from=5-1, to=5-2]
\end{tikzcd}\]
    \caption{Higgsing tree of $\mathbb F_5,\,\mathbb F_6,\,\mathbb F_7,\mathbb F_8,\,\mathbb F_{12},\,$}
    \label{fig:F5-12higgsingtree}
\end{subfigure}
\caption{$\mathbb F_2,\dots,\mathbb F_{12}$ Higgsing trees.}
\label{}
\end{figure}
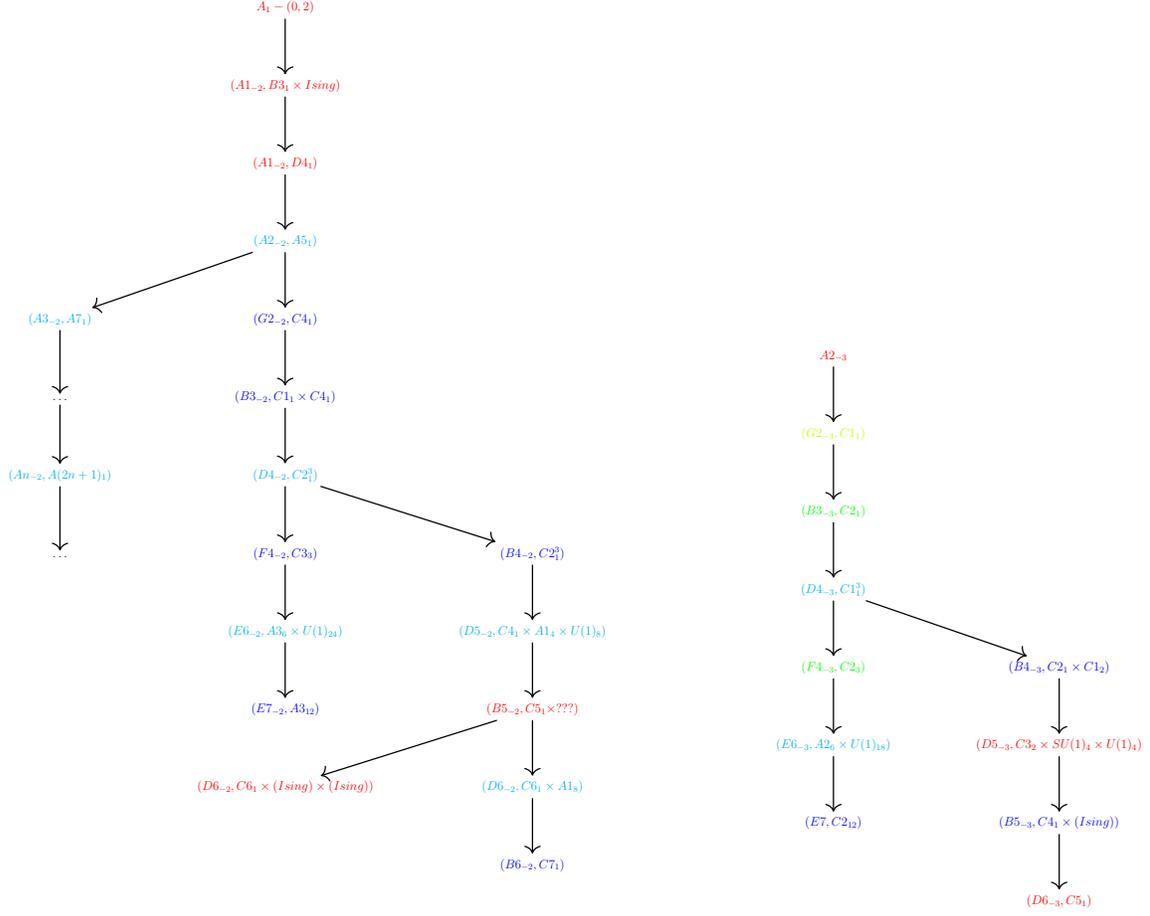
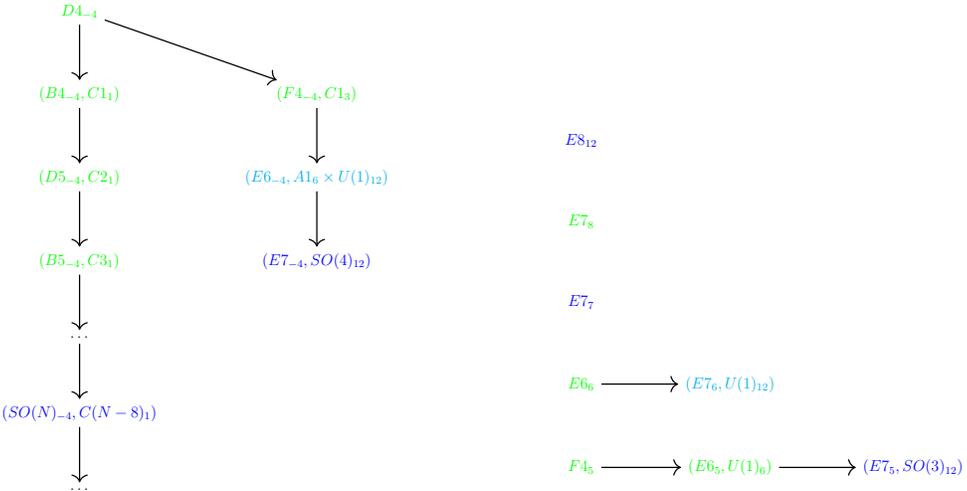